\documentclass[letterpaper,twocolumn,10pt]{article}
\usepackage{usenix2019}

\usepackage{epsfig,endnotes}
\usepackage{balance}
\usepackage{bold-extra}
\usepackage{syntax}
\usepackage{syntax}
\usepackage{color}
\usepackage{listings}
\usepackage{hyperref}
\usepackage{listings}
\usepackage{tikz}
\usepackage{xspace}
\usepackage{float}
\usepackage{placeins}
\usepackage[font=small,labelfont=bf]{caption}
\usepackage{subcaption}
\usepackage{wrapfig}
\usepackage{inconsolata}
\usepackage{url}
\usepackage{graphicx}
\usepackage{mdframed}
\usepackage[english]{babel}
\usepackage{blindtext}
\usepackage{stmaryrd}
\usepackage{eulervm}
\usepackage{booktabs}
\usepackage{mathtools}
\usepackage{amssymb}
\usepackage{amsthm}
\usepackage[capitalise]{cleveref}
\usepackage{algorithm}
\usepackage{algpseudocode}
\usepackage{tikz,pgf}

 \usetikzlibrary{circuits.logic.US}
 \usetikzlibrary{arrows}
\usepackage{pifont}
\usepackage{multirow}
\usepackage{xargs} 
\usepackage[colorinlistoftodos,prependcaption]{todonotes}

\usepackage{mathpartir}
\usepackage{lineno}

\usepackage{nopageno} 

\hypersetup{colorlinks=true,linkcolor=blue,urlcolor=blue}
\usepackage[compact,nonindentfirst]{titlesec}

\usepackage[normalem]{ulem}

\RequirePackage{commands}

\definecolor{vgreen}{RGB}{104,180,104}
\definecolor{vblue}{RGB}{49,49,255}
\definecolor{vorange}{RGB}{255,143,102}

\lstdefinestyle{verilog-style}
{
    language=Verilog,
    basicstyle=\footnotesize\ttfamily,
    keywordstyle=\color{vblue},
    identifierstyle=\color{black},
    commentstyle=\color{vgreen},
    numbers=left,
    numberstyle=\tiny\color{black},
    extendedchars=true,
    numbersep=10pt,
    tabsize=4,
    moredelim=*[s][\colorIndex]{[}{]},
    literate=*{:}{:}1,
    xleftmargin=5.0ex,
    captionpos=b,
    escapechar=\$
}

\makeatletter
\newcommand*\@lbracket{[}
\newcommand*\@rbracket{]}
\newcommand*\@colon{:}
\newcommand*\colorIndex{%
    \edef\@temp{\the\lst@token}%
    \ifx\@temp\@lbracket \color{black}%
    \else\ifx\@temp\@rbracket \color{black}%
    \else\ifx\@temp\@colon \color{black}%
    \else \color{vorange}%
    \fi\fi\fi
}
\makeatother

\crefname{lemma}{lemma}{lemmas}
\crefname{section}{\S}{\S}
\Crefname{section}{\S}{\S}

\newenvironment{CompactItemize}%
  {\begin{list}{$\blacktriangleright$}%
   {\leftmargin=\parindent \itemsep=2pt \topsep=2pt
     \parsep=0pt \partopsep=0pt}}%
  {\end{list}}

\begin{document}

\title{\sys: Verifying Constant-Time Execution of Hardware} 

\date{}

\author{
{\rm Klaus v. Gleissenthall}\\
University of California, San Diego
\and
{\rm Rami Gökhan Kıcı}\\
University of California, San Diego
\and
{\rm Deian Stefan}\\
University of California, San Diego
\and
{\rm Ranjit Jhala}\\
University of California, San Diego
} 

\maketitle

\thispagestyle{empty}

\mypara{Abstract}
To be secure, cryptographic algorithms crucially rely on the
underlying hardware to avoid inadvertent leakage of secrets through
timing side channels.
Unfortunately, such timing channels are ubiquitous in modern hardware,
due to its labyrinthine fast-paths and optimizations.
A promising way to avoid timing vulnerabilities is to devise---and
verify---conditions under which a hardware design is free of timing
variability, \ie executes in \emph{constant-time}.
%
%
In this paper, we present \sys: a clock-precise, constant-time
approach to eliminating timing side channels in hardware.
%
%
%
\sys succeeds in verifying various open source hardware designs in
seconds and with little developer effort.
\sys also discovered two constant-time violations: 
one in a floating-point unit and another 
one in an RSA encryption module.


\section{Introduction}
\label{sec:intro}
Trust in software systems is always rooted in the underlying hardware.
This trust is apparent when using hardware security features like
enclaves (\eg SGX and TrustZone), crypto units (\eg AES-NI and the
TPM), or MMUs.
But our trust goes deeper.
Even for simple \verb|ADD| or \verb|MUL| instructions, we expect the
processor to avoid leaking any of the operands via \emph{timing side
channels}, \eg by varying the execution time of the operation
according to the data.
Indeed, even algorithms specifically designed to be resilient to such
timing side-channel attacks crucially rely on these
assumptions~\cite{salsa20, poly1305, donnacurve}.
Alas, recently discovered vulnerabilities have shown that the
labyrinthine fast-paths and optimizations ubiquitous in modern
hardware expose a plethora of side channels that undermine many of our
deeply held beliefs~\cite{spectre, kohlbrenner2017, Lipp2018meltdown}.

A promising way to ensure that trust in hardware is properly earned is
to formally specify our expectations, and then, to \emph{verify}---through
mathematical proof---that the units used in
security critical contexts
do not exhibit
any timing variability, \ie are \emph{constant-time}.
For instance, by verifying that certain parts of 
an arithmetic logic unit (ALU) are constant-time, 
we can provide a foundation for implementing 
secure crypto algorithms in software~\cite{Barthe14, Barthe04, CTSoftware}.
Dually, if timing variability is unavoidable, \eg in SIMD or
floating-point units, making this variability \emph{explicit} can better
inform mechanisms that attempt to mitigate timing channels at the
software level \cite{MyersPLDI2012, rane2016, Andrysco18} in order to
avoid vulnerabilities due to gaps in the hardware-software
contract~\cite{andrysco2015,Andrysco18}.

In this paper, we introduce \sys: a clock-precise, constant-time
approach to eliminating timing side channels in hardware. 
Given a hardware circuit described in Verilog, a \emph{specification}
comprising a set of sources and sinks (\eg an FPU pipeline start and
end) and a set of usage assumptions (\eg no division is performed),
\sys allows developers to automatically synthesize \emph{proofs} which
ensure that the hardware runs in constant-time, \ie under the given
usage assumptions, the time taken to flow from source to sink, is
independent of operands, processor flags and interference by
concurrent computations.
%
 
 Using \sys, a crypto hardware designer can be certain that their
encryption core does not leak secret keys or messages by taking a
different number of cycles depending on the secret values.
Similarly, a CPU designer can guarantee that programs (\eg
cryptographic algorithms, SVG filters) will run in constant-time when
properly structured (\eg when they do not branch or access memory
depending on secrets~\cite{Barthe14}).

\sys is \emph{clock-precise} in that it 
enforces constant-time execution directly 
as a semantic property of the circuit rather 
than through indirect means like information 
flow control~\cite{SecVerilog}.
As a result, \sys neither requires the 
constant-time property to hold unconditionally 
nor demands the circuit be partitioned 
between different security levels 
(\eg as in SecVerilog~\cite{SecVerilog}). 
This makes \sys particularly suited for 
verifying existing hardware designs.
For example, we envision \sys to be useful in verifying ARM's recent
set of \emph{data independent timing (DIT)} instructions which should
execute in constant-time, if the \stmt{PSTATE.DIT} processor state
flag is set~\cite{linux-on-arm, arm-a64-isa}.

While there have been significant strides in verifying the
constant-time execution of software~\cite{CTSoftware, Barthe14,
Barthe-Product, Barthe04, watt:2019:ctwasm, jasmin,
verify-s2n-mee-cbc, Andrysco18}, \sys unfortunately cannot directly
reuse these efforts.
Constant time methods for software focus on
straight-line, sequential---often cryptographic---code.

Hardware designs, however, are inherently \emph{concurrent}
and \emph{long-lived}: circuits can be viewed as collections of
processes that run forever, performing parallel computations that
update registers and memory in every clock cycle. 
As a result, in hardware, even the definition of constant-time
execution becomes problematic: how can we measure the timing of a
hardware design that never stops and performs multiple concurrent
computations that mutually influence each other?

In \sys, we address these challenges through the following
contributions.

\mypara{1. Definition} 
First, we define a notion of constant-time execution for concurrent,
long-lived computations.  In order to reason about the timing of
values flowing between sources and sinks, we introduce the notion of
\emph{influence set}. The influence set of a value contains all
cycles~$t$, such that an input (\ie a source value) at~$t$ was used in
its computation. We say that a hardware design is constant time, if all
its computation paths (that satisfy usage assumptions) produce the
same sequence of influence sets for sinks.
%


\mypara{2. Verification}
To enable its efficient verification, we show how to reduce the
problem of checking constant-time execution---as defined through
influence sets---to the standard problem
of checking assertion validity.
For this, we first eschew the complexity of reasoning about several
concurrent computations at once, by focusing on a \emph{single} computation
starting (\ie inputs issued) at some cycle~$t$.  We say that a value
is \emph{live} for cycle~$t$ ($t$-live), if it was influenced by the
computation started at~$t$, \ie $t$ is in the value's influence set. This
allows us to reduce the problem of checking equality of influence
sets, to checking the equivalence of membership, for their elements.
We say that a hardware design is \emph{liveness equivalent},
if, for any two executions (that satisfy usage assumptions), and any
$t$, $t$-live values are assigned to sinks in the same way, \ie
whenever a $t$-live value is assigned to a sink in one execution, a
$t$-live value must also be assigned to a sink in the other. 

%
%
To check a hardware design for liveness equivalence, we
\emph{mark} source data as live in some \emph{arbitrarily
chosen} start cycle~$t$, and track the flow of $t$-live values through the circuit
using a simple standard taint tracking monitor~\cite{magazinius2010fly};
the problem of checking liveness equivalence then reduces
to checking a simple assertion stating that sinks are always tainted
in the same way.
Reducing constant-time execution to the standard problem of checking
assertion validity allows us to rely on off-the-shelf, mature
verification technology, which explains \sys{}'s effectiveness.

  

\mypara{3. Evaluation} 
Our final contribution is an implementation 
and evaluation of \sys on seven open source 
\Verilog projects---CPU cores, an ALU, crypto-cores, 
and floating-point units (FPUs).
We find that \sys succeeds in verifying different kinds of hardware
designs in a matter of seconds, with modest developer effort
(\S~\ref{sec:evaluation}).
Many of our benchmarks are constant-time for intricate reasons
(\cref{sec:case-studies}), \eg whether or not a circuit is
constant-time depends on its execution history, circuits are
constant-time despite triggering different control flow paths
depending on secrets, and require a carefully chosen set of
assumptions to be shown constant-time.
In our experience, these characteristics---combined with the circuit
size---make determining whether a
hardware design is constant-time by code inspection near
impossible. 

\sys also revealed two constant-time violations: one in the division unit
of an FPU designs, another in the modular exponentiation module of an RSA
encryption module. 
The second violation---a classical timing side channel---can be abused to leak
secret keys~\cite{kocher1996timing,brumley2005remote}.
%

In summary, this paper makes the following contributions.
\begin{CompactItemize}
\item First, we give a definition for 
      constant-time execution of hardware, 
      based on the notion of 
      \emph{influence sets}~(\cref{sec:overview}). 
      We formalize the semantics of \Verilog programs with influence
sets (\cref{sec:semantics}), and use this formalization to define
constant-time execution with respect to usage assumptions
(\cref{sec:taint}).
      \item Our second contribution is a reduction of constant-time
execution to the easy-to-verify problem of liveness equivalence.
      We formalize this property (\cref{sec:taint}), prove its
equivalence to our original notion of constant-time
execution~(\cref{equiv-proof}), and show how to verify it using
standard methods (\cref{sec:vcgen}).
      \item Our final contribution is an implementation and evaluation
of \sys on several challenging open source
hardware designs (\cref{sec:evaluation}). Our evaluation shows that
\sys can be used to verify constant-time execution of existing
hardware designs, rapidly, and with modest user effort.
\end{CompactItemize}


%
\section{Overview}
\label{sec:overview}

\begin{figure}[t]
  \begin{lstlisting}[style=verilog-style]
// source(x); source(y); sink(out);
// assume(ct = 1);

reg flp_res, x, y, ct, out, out_ready, ...;
wire iszero, isNaN, ...;

assign iszero = (x == 0) || (y == 0);

always @(posedge clk) begin
  ...
  flp_res <= ... // (2) compute x * y
end

always @(posedge clk) begin
  if (ct)
    ...; out <= flp_res; // (4)
  else
    if (iszero)
      out <= 0; // (1)
    else if (isNaN)
      ...
    else out <= flp_res; // (3)
    end
  end
end
  \end{lstlisting}
  \caption{Floating point multiplier (\exNum{1}).}
  \label{fig:mult-run}
\end{figure}

In this section, we give an overview of \sys and show how our tool can
be used to verify that a piece of \Verilog code executes in
constant-time.
As a running example, we consider a simple implementation of a
floating-point multiplication unit.
%

\mypara{Floating Point Multiplier} Our running example, like most FPUs, is generally \emph{not}
constant-time---common operations (\eg multiplication by zero) are
dramatically faster than rare ones (\eg multiplication by denormal
numbers~\cite{andrysco2015, kohlbrenner2017}).
But, like the ARM's recent support for data independent timing
instructions, our FPU
contains a processor flag that can be set to ensure that all multiplications
are constant-time, at the cost of performance.
\cref{fig:mult-run} gives a simplified fragment of \Verilog code that
implements this FPU multiplier.
While our benchmarks consist of hundreds of threads with shared
variables, pipelining, and contain a myriad of branches and flags
which cause dependencies on the execution history (see
\cref{sec:case-studies}), we have kept our running example as simple
as possible: our multiplier takes two floating-point values---input
registers $\stmt{x}$ and $\stmt{y}$---and stores the computation
result in output register~$\stmt{out}$.
%
Recall that \Verilog programs operate
on two kinds of data-structures: \emph{registers} which are assigned
in $\alwaysName$-blocks and store values across clock cycles and
\emph{wires} which are assigned in $\assignNameHL$-blocks and hold
values only within a cycle. 
Control register~$\stmt{ct}$ is used to configure the FPU to run in
constant-time (or not).
For simplicity, we omit most other control logic (\eg reset or output-ready
bits and processing of inputs).
Internally, the multiplier consists of several \emph{fast} paths and a single
\emph{slow} path.
For example, to implement multiplication by zero, one, $\NaN$, and other
special values we, inspect the input registers for these values and
produce a result in a single cycle (see ${\CommentColor (1)}$).
Multiplication by other numbers is more complex, however, and generally takes
more than a single cycle.
As shown in \cref{fig:mult-run}, this \emph{slow} path consists of
multiple intermediate steps, the final result of which is assigned to a
temporary register~$\stmt{flp\_res}$ (see ${\CommentColor (2)}$) before
$\stmt{out}$ (see ${\CommentColor (3)}$).\footnote{
  For simplicity, we omit the intermediate steps and assume that they implement
  floating-point multiplication in constant-time. In practice, FPUs may also take 
  different amounts of time depending on such values.
}
Importantly, when the constant-time configuration register~$\stmt{ct}$ is
set, only this slow path is taken (see ${\CommentColor (4)}$).

\mypara{Outline} In the rest of this section, we show how \sys verifies that this
hardware design runs in constant-time when the $\stmt{ct}$ flag is set
and violates the constant-time property otherwise.
We present our definition of constant-time based on of influence sets
in \cref{ct-def}, liveness equivalence in \cref{ct-live}, and
finally show how \sys formally verifies liveness equivalence by
reducing it to a simple safety property that can be handled by
standard verification methods \cref{ct-spec}.

\begin{figure}[t]
  \figuresize
  \centering
  \begin{equation*}
    \begin{array}[t]{@{}r@{\;\;\;}c@{\;\;}l@{\qquad}l@{}}
  P & ::= && \textbf{Program}\\[1ex]
      && | \;\; \proc{\statement}_{\mathit{id}} & \textit{process} \\
      && | \;\; P \inParSync P & \textit{parallel composition} \\
       && | \;\; \srepeat{P} & \textit{sync. iteration} \\
       && | \;\; \emp & \textit{empty process} \\
      
     \statement & ::= && \textbf{Command}\\
             && | \;\; \sskip  & \textit{no-op} \\
       && | \;\; \syncAssign{v}{e}  & \textit{blocking }\\
       && | \;\; \asyncAssign{v}{e}  & \textit{non-blocking
                                       }\\
       && | \;\; \contInter{v}{e}  & \textit{continuous} \\
       && | \;\; \ite{e}{\statement}{\statement} & \textit{conditional} \\
    && | \;\; \statement_1 \mathrel{;} \dots \mathrel{;}
          \statement_k & \textit{sequence} \\
      && | \;\; a & \textit{annotation} \\
      e & ::= && \textbf{Expression} \\[\jot]
&&| \;\; v & \textit{variables} \\
&&| \;\; n & \textit{constants} \\
      &&| \;\; \func{e_1, \dots, e_k} & \textit{function literal}
    \end{array}
  \end{equation*}
  \caption{Syntax for intermediate language \InterLang.}
  \label{fig:synclang-syntax}
\end{figure}

\subsection{Constant-Time For Hardware \label{ct-def}} 

We start by defining \emph{constant-time} execution for hardware.

\mypara{Assumptions and Attacker Model}
Like SecVerilog~\cite{SecVerilog}, we scope our work to synchronous circuits
with a single, fixed-rate clock.
We further assume an \emph{external attacker} that can measure the execution
time of a piece of hardware (given as influence sets) using a
cycle-precise timer. In particular, an attacker can observe the timing of
 \emph{all} inputs that influenced a given output.
%
%
These assumptions afford us many benefits.
(Though, as we describe in \Cref{sec:limitations}, they are not for free.)
For example, assuming a single fixed-rate clock, allows us to
translate \Verilog programs, such as our FPU multiplier to a more
concise representation shown in \Cref{ex:running}.
 
\begin{figure}
\figuresize
\[
  \begin{array}[c]{@{}r@{\quad}l@{\quad}c@{}l@{}}
     &

       \repeatName &
       [
    \contInter{ \stmt{iszero} } 
    { \left (
      \binOr{\stmt{x} \doubleEqual 0}{\stmt{y}\doubleEqual 0 } \right )  
    }
     & ] \\[\jot]
    \mathrel{\inParSync} &
                           \repeatName & \proc{\dots \mathrel{;} \asyncAssign{\stmt{flp\_res}} { \dots}
    }  \\[2ex]
    \mathrel{\inParSync}
            &
      
    \repeatName &  
              \left [   
              \;\;
              \begin{array}[c]{@{}l@{}}
                                  \iteName ( \stmt{ct}, \\
              \quad \asyncAssign{\stmt{out}}{\stmt{flp\_res}}, \\
              \quad \iteName ( \stmt{iszero}, \\
                \quad \quad\asyncAssign{\stmt{out}}{0}, \\   
                 \quad \quad
                \asyncAssign{\stmt{out}}{\stmt{flp\_res}}  ) )
              \end{array}
    \right ]
  \end{array}
\]
  \caption{\exNum{1} written in \InterLang}
  \label{ex:running}
\end{figure}
 
 \mypara{Intermediate Language} In this language---called
\InterLang---\Verilog $\alwaysName$- and $\assignNameHL$-blocks are
represented as concurrent \emph{processes}, wrapped inside an infinite
$\repeatName$-loop.
As \cref{fig:synclang-syntax} shows, each process sequentially
executes a series of \Verilog-like statements.
(Each process also has a unique identifier $\ID \in \ProcIDs$, which we sometimes
omit, for brevity.)
Most of these are standard; we only note that \InterLang---like
\Verilog---supports three types of assignment statements: blocking
($\syncAssign{v}{e}$), non-blocking ($\asyncAssign{v}{e}$) and
continuous ($\contInter{v}{e}$).
Blocking assignments take effect immediately, within the current cycle;
non-blocking assignments are deferred until the next cycle. 
Finally, continuous assignments enforce directed equalities between
registers or wires: whenever the
right-hand side of an equality is changed, the left-hand side is updated
by re-running the assignment. Note that \InterLang focuses only on the
synthesizable fragment of \Verilog, \ie does not model delays,
etc., which are only relevant for simulation.

\vinter processes are composed in parallel using the $(\inPar)$~operator.
Unlike concurrent software processes, they are, however, synchronized
using a single (implicit) fixed-rate clock:
each process waits for all other (parallel) processes to finish executing
before moving on to the next iteration, \ie next clock
cycle. Moreover, unlike software, these programs are usually data-race
free, in order to be synthesizable to hardware.

\vinter processes run forever; they perform computations and update registers
(\eg $\stmt{out}$ in our multiplier) on every clock cycle.
For example, pipelined hardware units execute multiple, different
computations simultaneously.

\mypara{From Software to Hardware}
This execution model, together with the fact that software operates at
a higher level of abstraction than hardware, makes it difficult for us
to use existing verification tools for constant-time software
(\eg~\cite{CTSoftware, Barthe14}).

First, constant-time verification for software only considers
straight-line, sequential code.
This makes it ill-suited for the concurrent, long-lived execution model
of hardware.
%

Second, software constant-time models are necessarily conservative.
They deliberately abstract over hardware details---\ie they don't
rely on a precise hardware models (\eg of caches or branch
predictors)---and instead use \emph{leakage models} that make control
flow and memory access patterns observable to the attacker.
This makes constant-time software portable across hardware. But, it
also makes the programming model restrictive: the model disallows any
branching to protect against hidden microarchitectural state (\eg the branch
predictor).
%


Since we operate on \Verilog, where all state is explicit and visible, we can
instead directly track the influence of secret values on the timing of
attacker-observable outputs.
This allows us to be more permissive than software constant-time
models.
For instance, if we can show that the execution of two branches of a hardware
design takes the same amount of time, independent of secret inputs, we can
safely allow branches on secrets. 
However, this still leaves the problem of pipelining: hardware ingests
inputs and produce outputs at every clock cycle: how then do we know
(if and) which secret inputs influenced a particular output?
%

%

\begin{figure*}
\figuresize
  \centering
  \begin{tabular}[h]{|c||c|c|c|c|c||c|c|c|c|c|}
    \hline
    Cycle \# & $\stmt{x}$       & $\stmt{y}$       & $\stmt{ct}$         & $\stmt{fr}$                  & $\stmt{out}$     & $\influenceOf{\stmt{x}}$ & $\influenceOf{\stmt{y}}$ & $\influenceOf{\stmt{ct}}$ & $\influenceOf{\stmt{fr}}$ & $\influenceOf{\stmt{out}}$ \\
    \hline
    0       & \cellhighlight $0$              & $1$              & \False              & $\UnknownValue$              & $\UnknownValue$  & $\theSet{0}$             & $\theSet{0}$             & $\emptyset$               & $\emptyset$               & $\emptyset$                \\
    \hline
    1       & \makeShaded{$0$} & \makeShaded{$1$} & \makeShaded{\False} & \makeShaded{$\UnknownValue$} & \cellhighlight $0$              & $\theSet{1}$             & $\theSet{1}$             & \makeShaded{$\emptyset$}  & \makeShaded{$\emptyset$}  & $\theSet{0}$               \\
    \hline
    \multicolumn{11}{|c|}{\vdots}\\ 
    \hline
    k-1     & \makeShaded{$0$} & \makeShaded{$1$} & \makeShaded{\False} & \cellhighlight $0$                          & \makeShaded{$0$} & $\theSet{k-1}$           & $\theSet{k-1}$           & \makeShaded{$\emptyset$}  & $\theSet{0}$              & $\theSet{k-2}$             \\
    \hline
    k       & \makeShaded{$0$} & \makeShaded{$1$} &
                                                    \makeShaded{\False} & \makeShaded{$0$}             & \cellhighlight \makeShaded{$0$} & $\theSet{k}$             & $\theSet{k}$             & \makeShaded{$\emptyset$}  & $\theSet{1}$              & \cellhighlight $\mathbf{\theSet{k-1}}$    \\
    \hline
  \end{tabular}
  \caption{
    Execution of $\exNum{1}$, where $\stmt{x}=0$ and $\stmt{y}=1$, and $\stmt{ct}$ is unset.
    For each variable and cycle, we show its current value and influence set.
    We assume that it takes~$k$ cycles to compute the output along the slow path,
    and abbreviate~$\stmt{flp\_res}$ as $\stmt{fr}$.
    $\UnknownValue$ denotes an unknown/irrelevant value. 
    Register \stmt{out} is only influenced by values from the last cycle.
    Highlighted cells are the difference with \Cref{fig:non-ct-run-2-set}. Values that stayed the same in the next cycle are shaded. }
  \label{fig:set-non-ct-run-1}
\end{figure*}

\begin{figure*}
  \figuresize
  \centering
  \begin{tabular}[h]{|c||c|c|c|c|c||c|c|c|c|c|}
    \hline
     Cycle \# & $\stmt{x}$       & $\stmt{y}$       & $\stmt{ct}$         & $\stmt{fr}$                  & $\stmt{out}$                 & $\influenceOf{\stmt{x}}$ & $\influenceOf{\stmt{y}}$ & $\influenceOf{\stmt{ct}}$ & $\influenceOf{\stmt{fr}}$ & $\influenceOf{\stmt{out}}$ \\
    \hline
    0       & \cellhighlight $1$              & $1$              & \False              & $\UnknownValue$              & $\UnknownValue$              & $\theSet{0}$             & $\theSet{0}$             & $\emptyset$               & $\emptyset$               & $\emptyset$                \\
    \hline
    1       & \makeShaded{$1$} & \makeShaded{$1$} & \makeShaded{\False} & \makeShaded{$\UnknownValue$} & \cellhighlight{$\UnknownValue$} & $\theSet{1}$             & $\theSet{1}$             & \makeShaded{$\emptyset$}  & \makeShaded{$\emptyset$}  & $\theSet{0}$               \\
    \hline
    \multicolumn{11}{|c|}{\vdots}\\ 
    \hline
    k-1     & \makeShaded{$1$} & \makeShaded{$1$} & \makeShaded{\False} & \cellhighlight $1$                          & \makeShaded{$\UnknownValue$} & $\theSet{k-1}$           & $\theSet{k-1}$           & \makeShaded{$\emptyset$}  & $\theSet{0}$              & $\theSet{k-2}$             \\
    \hline
    k       & \makeShaded{$1$} & \makeShaded{$1$} & \makeShaded{\False} & \makeShaded{$1$}             & \cellhighlight $1$                          & $\theSet{k}$             & $\theSet{k}$             & \makeShaded{$\emptyset$}  & $\theSet{1}$              & \cellhighlight $\mathbf{\theSet{0, k-1}}$ \\ 
    \hline
  \end{tabular}
  \caption{
    Execution of $\exNum{1}$, where both $\stmt{x}=1$ and $\stmt{y}=1$, and $\stmt{ct}$ is unset.
    The execution produces the same influence sets as the execution in
    \cref{fig:set-non-ct-run-1}, except for cycle~$k$, where $\stmt{out}$'s
    influence set contains the additional value~$0$, thereby violating our
    definition of constant-time execution.
  }
  \label{fig:non-ct-run-2-set}
\end{figure*}

\mypara{Influence Sets}
This motivates our definition for \emph{influence sets}.
In order to define a notion of constant-time execution that is
suitable for hardware, we first add annotations marking inputs (\ie
$\stmt{x}$ and $\stmt{y}$ in our example) as \emph{sources} and
outputs (\ie $\stmt{out}$) as \emph{sinks}.
For a given cycle, we then associate with each register~$x$ its
\emph{influence-set} $\influenceOf{x}$. The influence set of a
register~$x$ contains all cycles~$t$, such that an input
at~$t$ was used in the computation of~$x$'s current value.
This allows us to define constant-time execution for hardware: we say
that a hardware design is constant-time, if any two executions (that
satisfy usage assumptions) produce the same sequence of influence sets
for their sinks.

\mypara{Example} We now illustrate this definition using our running
example~$\exNum{1}$ by showing that~$\exNum{1}$
violates our definition of constant-time, if the $\stmt{ct}$ flag is
unset.
For this, consider \cref{fig:set-non-ct-run-1}
and~\cref{fig:non-ct-run-2-set}, which show the state of registers and
wires as well as their respective influence sets, for two executions. In both
executions, we let $\stmt{y}=1$, but vary the value of the $\stmt{x}$
register: in \cref{fig:set-non-ct-run-1}, we set $\stmt{x}$ to $0$ to
trigger the fast path in \cref{fig:non-ct-run-2-set} we set it
to~$1$.
In both executions, sources $\stmt{x}$ and $\stmt{y}$ are only
influenced by the current cycle, constant-time flag~$\stmt{ct}$ is set
independently of inputs, and temporary register $\stmt{flp\_res}$ is
influenced by the inputs that were issued $k-1$ cycles ago, as it takes
$k-1$ cycles to compute $\stmt{flp\_res}$
along the slow path.

The two executions differ in the influence sets of~$\stmt{out}$. In
\cref{fig:set-non-ct-run-1}, \stmt{out} is only influenced by the input
issued in the last cycle, through a control dependency on
$\stmt{iszero}$. In the execution in \cref{fig:non-ct-run-2-set}, its
value at cycle $k$ is however also
influenced by the input at $0$.
This reflects the propagation of the computation result through the
slow path.
Crucially, it also shows that the multiplier is not
constant-time---the sets $\influenceOf{\stmt{out}}$ differing between
two runs reflects the influence of data on the duration of the
computation.

\subsection{ Liveness Equivalence \label{ct-live}}
We now show how to reduce verifying whether a given
hardware is constant-time to an easy-to-check, yet equivalent problem
called liveness equivalence. Intuitively, liveness equivalence reduces
the problem of checking equality of influence sets, to checking the
equivalence of membership, for arbitrary elements.

\mypara{Liveness Equivalence}
Our reduction focuses on a single computation started at some
cycle~$t$.
We say that register~$x$ is \emph{live} for cycle $t$ ($t$-live), if
its current value is influenced by an input issued in cycle $t$, \ie
if $t \in \influenceOf{x}$.
Two executions are $t$-liveness equivalent, if whenever a $t$-live
value is assigned to a sink in one execution, a $t$-live value must
also be assigned in the other.
Finally, a hardware design is liveness equivalent, if any two
executions that satisfy usage assumptions are~$t$-liveness equivalent,
for any~$t$.

\mypara{Live Value Propagation}
To track $t$-liveness for a fixed~$t$, \sys internally transforms
\vinter programs as follows.
For each register or wire (\eg $\stmt{x}$ in our multiplier), we introduce a
new shadow variable (\eg $\taintOf{\stmt{x}}$) that represents its
liveness;
a shadow variable $\taintOf{\stmt{x}}$ is set to $\TTrue$ if \stmt{x} is live
and $\TFalse$ (dead) otherwise.\footnote{
For liveness-bits $\taintOf{\stmt{x}}$ and $\taintOf{\stmt{y}}$, we define a join
operator~$\lor$, such that $\taintOf{\stmt{x}}\lor\taintOf{\stmt{y}}$
is $\TTrue$, if $\taintOf{\stmt{x}}$ or $\taintOf{\stmt{y}}$
is~$\TTrue$ and $\TFalse$, otherwise.
}
We then propagate liveness using a standard taint-tracking inline
monitor~\cite{magazinius2010fly} shown in \Cref{eq:progtaint}.
\begin{figure}
  \figuresize
  \[
  \begin{array}[t]{@{}r@{\quad}l@{\quad}c@{\quad}r@{}}

  &
\repeatName & \left  [ \;
              \begin{array}[c]{@{}l@{}}
       \contInter{ \stmt{iszero} } 
       { \left ( 
       \binOr{\stmt{x}\doubleEqual 0} {\stmt{y} \doubleEqual 0}\right )  
       }
       \mathrel{;} \\                   
       \contInter{\taintOf{\stmt{iszero}}}                   
       { \left ( 
       \taintOf{\stmt{x}}\lor \taintOf{\stmt{y}}
       \right )  }
   \end{array}   
  \;  \right ] & 
    \\[3ex]                  
     \mathrel{\inPar}  & 
     
   \repeatName &\left  [  \; 
     \begin{array}[c]{@{}l@{}}
       \dots \mathrel{;} \asyncAssign{\stmt{flp\_res}} {\dots } 
       \mathrel{;} \\
       \dots \mathrel{;} \asyncAssign{\taintOf{\stmt{flp\_res}}}
       { \dots  {\CommentColor \; \sComment
       \left ( 
    \taintOf{\stmt{x}}
    \lor 
    \taintOf{\stmt{y}}
       \right )
       }
       }
 \end{array}                     
  \;   \right ] &
    \\[3ex]

    \mathrel{\inPar}
    &
      \repeatName   
      & \left [ \;
      \begin{array}[c]{@{}l@{}}
        \iteName(\stmt{ct}, \\
        \quad 
        \begin{array}[l]{@{}l@{}}
          \asyncAssign{\stmt{out}}{\stmt{flp\_res}} \mathrel{;}  \\
          \asyncAssign{\taintOf{\stmt{out}}} {
          \left (
          \taintOf{\stmt{flp\_res}} \lor
          \taintOf{\stmt{ct}}
          \right )
          } ,
        \end{array} \\ 
        \quad  
        \iteName(\stmt{iszero}, \asyncAssign{\stmt{out}}{0} \mathrel{;} \\ 
        \quad \quad
        \begin{array}[l]{@{}l@{}}
          \asyncAssign{\taintOf{\stmt{out}}}{ 
          \left (  \taintOf{\stmt{ct}} \lor
          \taintOf{\stmt{iszero}} \right )} , \\
          \asyncAssign{\stmt{out}}{\stmt{flp\_res}} \mathrel{;} \\
                      \asyncAssign{\taintOf{\stmt{out}}} {
          \left (
          \begin{array}[c]{@{}c@{}}
            \taintOf{\stmt{flp\_res}} \lor \\
            \taintOf{\stmt{ct}} \lor 
            \taintOf{\stmt{iszero}} 
          \end{array}
          \right ) 
          } 
        ) )
        \end{array}
       %
      \end{array}
    \; \right ]
      & 
  \end{array}
  \]
  \caption{\exNum{1}, after we propagate liveness using a standard taint-tracking
    inline monitor.}
  \label{eq:progtaint}
\end{figure}
Intuitively, our monitor ensures that registers and wires that depend on a
live value---directly or indirectly, via control flow---are marked live.

\mypara{Example} By tracking liveness, we can again see that our
floating-point multiplier is not constant-time when the
$\stmt{ct}$ flag is unset.
To this end, we ``inject'' live values at sources ($\stmt{x}$ and
$\stmt{y}$) at time $t=0$ for two runs; as before, we set $\stmt{y}=1$,
and vary the value of $\stmt{x}$: in one execution, we set $\stmt{x}$ to $0$ to trigger the fast path,
in the other execution, we set it to~$1$.
\cref{fig:non-ct-run-1} and~\ref{fig:non-ct-run-2} show the state of the
different registers and wires for these runs.
In both runs, $\stmt{out}$ is live at cycle $1$---due to a control
dependency in \cref{fig:non-ct-run-1}, due a direct assignment in
\cref{fig:non-ct-run-2}.
But, in the latter, $\stmt{out}$ is \emph{also} live at the $k$th
cycle.
This reflects the fact that the influence sets of $\stmt{out}$ at
cycle~$k$ differ in the membership of $0$, and therefore witnesses the
constant-time violation.

\begin{figure}
\figuresize
  \centering
  \resizebox{\columnwidth}{!}{
  \begin{tabular}[h]{|c||c|c|c|c|c||c|c|c|c|c|}
    \hline
        & $\stmt{x}$         & $\stmt{y}$       & $\stmt{ct}$         & $\stmt{fr}$                  & $\stmt{out}$                     & $\taintOf{\stmt{x}}$ & $\taintOf{\stmt{y}}$ & $\taintOf{\stmt{ct}}$ & $\taintOf{\stmt{fr}}$ & $\taintOf{\stmt{out}}$ \\
    \hline
    0   & \cellhighlight $0$ & $1$              & \False              & $\UnknownValue$              & $\UnknownValue$                  & \TTrue               & \TTrue               & \TFalse               & \TFalse               & \TFalse                \\
    \hline
    1   & \makeShaded{$0$}   & \makeShaded{$1$} & \makeShaded{\False} & \makeShaded{$\UnknownValue$} & \cellhighlight $0$               & \TFalse              & \TFalse              & \makeShaded{\TFalse}  & \makeShaded{\TFalse}  & \TTrue                 \\
    \hline
    \multicolumn{11}{|c|}{\vdots}                                                                                                                                                                                                                                \\
    \hline
    k-1 & \makeShaded{$0$}   & \makeShaded{$1$} & \makeShaded{\False} & \cellhighlight $0$           & \makeShaded{$0$}                 & \makeShaded{\TFalse} & \makeShaded{\TFalse} & \makeShaded{\TFalse}  & \TTrue                & \TFalse                \\
    \hline
    k   & \makeShaded{$0$}   & \makeShaded{$1$} & \makeShaded{\False} & \makeShaded{$0$}             & \cellhighlight  \makeShaded{$0$} & \makeShaded{\TFalse} & \makeShaded{\TFalse} & \makeShaded{\TFalse}  & \TFalse               & \cellhighlight \TFalse                \\
    \hline
  \end{tabular}
  }
\caption{Execution of $\taintOf{\exNum{1}}$, where  $\stmt{x}=0$
  and $\stmt{y}=1$. We show current value and liveness bit for
  each register and cycle. Register $\stmt{out}$ is live in cycle one, due to
  the fast path and dead, otherwise. Highlights are the differences with \Cref{fig:non-ct-run-2}. Values that stayed the same in the next cycle are shaded.}
 \label{fig:non-ct-run-1}
\end{figure}

\begin{figure}
  \figuresize
  \centering
  \resizebox{\columnwidth}{!}{
  \begin{tabular}[h]{|c||c|c|c|c|c||c|c|c|c|c|}
    \hline
        & $\stmt{x}$         & $\stmt{y}$       & $\stmt{ct}$         & $\stmt{fr}$                  & $\stmt{out}$                    & $\taintOf{\stmt{x}}$ & $\taintOf{\stmt{y}}$ & $\taintOf{\stmt{ct}}$ & $\taintOf{\stmt{fr}}$ & $\taintOf{\stmt{out}}$ \\
    \hline
    0   & \cellhighlight $1$ & $1$              & \False              & $\UnknownValue$              & $\UnknownValue$                 & \TTrue               & \TTrue               & \TFalse               & \TFalse               & \TFalse                \\
    \hline
    1   & \makeShaded{$1$}   & \makeShaded{$1$} & \makeShaded{\False} & \makeShaded{$\UnknownValue$} & \cellhighlight  $\UnknownValue$ & \TFalse              & \TFalse              & \makeShaded{\TFalse}  & \makeShaded{\TFalse}  & \TTrue                 \\
    \hline
    \multicolumn{11}{|c|}{\vdots}                                                                                                                                                                                                                               \\
    \hline
    k-1 & \makeShaded{$1$}   & \makeShaded{$1$} & \makeShaded{\False} & \cellhighlight $1$           & \makeShaded{$\UnknownValue$}    & \makeShaded{\TFalse} & \makeShaded{\TFalse} & \makeShaded{\TFalse}  & \TTrue                & \TFalse                \\
    \hline
    k   & \makeShaded{$1$}   & \makeShaded{$1$} & \makeShaded{\False} & \makeShaded{$1$}             & \cellhighlight  $1$             & \makeShaded{\TFalse} & \makeShaded{\TFalse} & \makeShaded{\TFalse}  & \TFalse               & \cellhighlight \TTrue                 \\
    \hline
  \end{tabular}
  }
\caption{Execution of $\taintOf{\exNum{1}}$, where both $\stmt{x}=1$
and $\stmt{y}=1$. The
  liveness bits are the same as in~\ref{fig:non-ct-run-1}, except for
  cycle~$k$, where $\stmt{out}$ is now live. This reflects the
  propagation of the output value through the slow path and shows
  the constant-time violation.}
\label{fig:non-ct-run-2}
\end{figure}

\subsection{Verifying Liveness Equivalence \label{ct-spec}}
Using our reduction to liveness equivalence, we can \emph{verify} that
a \Verilog program executes in constant-time using standard methods.
For this, we \emph{mark} source data as live in some \emph{arbitrarily
chosen} start cycle~$t$. We then verify that any \emph{two} executions
that satisfy usage assumptions assign $t$-live values to sinks, in the
same way.

\mypara{Product Programs}
Like previous work on verifying constant-time
software~\cite{CTSoftware}, \sys reduced the problem of
verifying properties of \emph{two} executions of some program $P$ by
proving a property about a \emph{single} execution of a new program
$Q$.
This program---the so-called \emph{product program}~\cite{Barthe04}
-- consists of two disjoint copies of the original program.

\mypara{Race-Freedom}
 Our product construction exploits the fact that\Verilog programs are
 \emph{race-free}, \ie the order in which $\alwaysName$-blocks are
 scheduled within a cycle does not matter.
While races in software often serve a purpose (\eg a task distribution
service may allow races between equivalent worker threads to increase
throughput), races in \Verilog are always artifacts of poorly designed code:
any synthesized circuit is, by its nature, race-free, \ie the
scheduling of processes \emph{within} a cycle does not affect the
computation outcome.
Indeed, races in \Verilog represent an under-specification of the intended
design.

\mypara{Per-Process Product}
We leverage this insight to compose the two copies of a program in
\emph{lock-step}.
Specifically, we merge each process of the two program copies and execute the
``left'' (L) and ``right'' (R) copies together.
For example, \sys transforms the \vinter multiplier code from
\Cref{eq:progtaint} into the \emph{per-process product program} shown in
\Cref{fig:per-process-product}.
 
\begin{figure}
  \figuresize
  \[
  \begin{array}[t]{@{}r@{\;\;\;}l@{\;}c@{}}
    &
    \srepeat &
    \left  [
     \begin{array}[c]{@{}c@{}}
                          
    \contInter{\leftOf{\stmt{iszero}}}{ 
     \left (          
      \binOr{\leftOf{\stmt{x}} \doubleEqual 0} {\leftOf{\stmt{y}} \doubleEqual 0}\right )
                        } \mathrel{;}  \\
     \contInter{\rightOf{\stmt{iszero}}} {\left (
      \binOr{\rightOf{\stmt{x}}\doubleEqual 0} {\rightOf{\stmt{y}}\doubleEqual 0}\right )
       }
       \mathrel{;}  \\
       \contInter{\taintOf{\leftOf{\stmt{iszero}}}}                   
       { \left ( 
       \taintOf{\leftOf{\stmt{x}}}\lor \taintOf{\leftOf{\stmt{y}}}
       \right )  }
        \mathrel{;} \\
         \contInter{\taintOf{\rightOf{\stmt{iszero}}}}                   
       { \left ( 
       \taintOf{\rightOf{\stmt{x}}}\lor \taintOf{\rightOf{\stmt{y}}}
       \right )  }
    \end{array}       
    \right ] \\[6ex]

     \mathrel{\inParSync} &
       \srepeat &
       \left [
        \begin{array}[c]{@{}c@{}}
                              
          \dots \mathrel{;} \asyncAssign{\leftOf{\stmt{flp\_res}}} {
          \dots 
          } \mathrel{;} \\
          \dots \mathrel{;} \asyncAssign{\rightOf{\stmt{flp\_res}}} {
          \dots
          } \mathrel{;} \\
          \asyncAssign{\taintOf{\leftOf{\stmt{flp\_res}}}}
       {
         \dots { \CommentColor \; \sComment
       \left ( 
    \taintOf{\leftOf{\stmt{x}}}
    \lor 
    \taintOf{\leftOf{\stmt{y}}}
       \right )
          }
       } \mathrel{;} \\
          \asyncAssign{\taintOf{\rightOf{\stmt{flp\_res}}}}
       {
          \dots {\CommentColor \; \sComment  
       \left ( 
    \taintOf{\rightOf{\stmt{x}}}
    \lor 
    \taintOf{\rightOf{\stmt{y}}}
       \right )
          }
       }
        \end{array}
    \right ]
    \\[\jot]
    \mathrel{\inParSync}
    &  \srepeat &
      \dots
  \end{array}
  \]
  \caption{Per-process product form of \exNum{1}.}
  \label{fig:per-process-product}
\end{figure}
 
Merging two copies of a program as such is sound:
since the program is race-free---any ordering of process transitions
\emph{within} a cycle yields the same results---we are
free to pick an arbitrary schedule.\footnote{To ensure that hardware designs
are indeed race-free, our implementation performs a light-weight static
analysis to check for races.}
Hence, \sys takes a simple ordering approach and schedules the left and right
copy of same process at the same time.

\mypara{Constant-Time Assertion}
Given such a product program, we can now frame the constant-time verification
challenge as a simple \emph{assertion}: the liveness of the left and right
program sink-variables must be the same (regardless of when the computation
started). 
In our example, this assertion is simply
$\leftOf{\taintOf{\stmt{out}}}=\rightOf{\taintOf{\stmt{out}}}$.
This assertion can be verified using standard methods. In particular,
\sys synthesize process-modular invariants~\cite{OwickiGries}
that imply the constant-time assertion (\cref{sec:algorithm}).

The following two sections formalize the material presented in this overview.
\section{Syntax and Semantics}
\label{sec:semantics}
Since \Verilog{}'s execution model can be
subtle~\cite{Verilog-standard}, we formally define syntax and semantics of the
\Verilog fragment considered in this paper.
\subsection{Preliminaries}
%
For a function~$f$, we write $\domain \; f$ to
denote~$f$'s domain and $\range \; f$ for its co-domain.
%
%
For a set~$S \subseteq \domain \; f$, we
let~$\upd{f}{S}{b}$ denote the function that behaves the same as~$f$
except $S$, where it returns~$b$, \ie $\upd{f}{S}{b}(x)$ evaluates
to~$b$ if $x\in S$ and $f(x)$, otherwise.
We use $\upd{f}{a}{b}$ as a short hand for $\upd{f}{\theSet{a}}{b}$.   
Sometimes,
we want to update a function by setting the function values of some
subset $S$ of its domain to a non-deterministically chosen value. For $S
\subseteq \domain \; f$, we write $\upd{f}{S}{\ast}(x)$ to denote the
function that evaluates to some $y$ with $y \in \range \; f$, if $x\in
S$ and $f(x)$ otherwise.
%
%
\subsection{Syntax}
%



%

We restrict ourselves to the \emph{synthesizable} fragment of
\Verilog, \ie we do not include commands like initial blocks that only
affect simulation and implement a \emph{normalization
step}~\cite{gordon95} in which the program is ``flattened'' by
removing module instantiation through in-lining. 
We provide \Verilog syntax and a translation to \InterLang in
\Cref{sec:syntax-trans}, but define semantics in \InterLang
(\cref{fig:synclang-syntax}).

%

\mypara{Annotations}
We define annotations in Figure~\ref{fig:annot-syntax}.
Let $\Register$ denote the set of registers and $\Wires$ the set of
wires and let~$\Vars$ denote their disjoint union, \ie $\Vars \eqdef
\Register \uplus \Wires$.
For a register~$v \in \Register$, annotations $\source{v}$ and $\sink{v}$
designate $v$ as source or sink, respectively.\footnote{To use
  wires as source/sink, one has to define an auxiliary register.} 
We let $\SourceSink \eqdef (\Sources, \Sinks)$ denote the set of
input/output assumptions, where~\Sources denotes the set of all
sources and \Sinks denote the set of all sinks.
Let $\varphi$  be a first-order formula over some background
theory that refers to two disjoint sets of variables~$\leftOf{Vars}$
and $\rightOf{\Vars}$.
Then, annotations~$\initEqual{\varphi}$ and $\alwaysEqual{\varphi}$ indicate
that formula~$\varphi$ holds initially or throughout the execution.
The assumptions are collected in $\Assumptions \eqdef \left (
\IntitEqVars, \AlwaysEqVars \right )$, such that $\IntitEqVars$ contains
all formulas under $\initEqualName$ and
$\AlwaysEqVars$ all formulas under $\alwaysEqualName$.

\begin{figure}[t]
  \centering
  \footnotesize
  \begin{equation*}
    \begin{array}[t]{@{}r@{\;}c@{\;}l@{\;\;}l@{\quad\quad}l@{\;\;}l@{}}
      a & \scriptstyle{::=} &                   & \textbf{In/Out} &                              & \textbf{Assump.}                 \\
        &                   & | \;\; \source{v} & \textit{source} & | \;\; \initEqual{\varphi}   & \textit{initially} \; \varphi \; \\
        &                   & | \;\; \sink{v}   & \textit{sink}   & | \;\; \alwaysEqual{\varphi} & \textit{always} \; \varphi
    \end{array}
  \end{equation*}
  \caption{Annotation syntax.}
  \label{fig:annot-syntax}
\end{figure}

\begin{figure}[t]
  \centering
  \footnotesize
  \begin{tabular}[t]{@{}cl@{\hskip 1cm}cl@{}}
    \toprule
    \textbf{Config} & \textbf{Meaning}       & \textbf{Trace}            & \textbf{Meaning}     \\
    \midrule
    \store          & \emph{store}           & $\Sigma$                  & \emph{configuration} \\
    \taintMap       & \emph{liveness map}    & $l$                       & \emph{label}         \\
    \influenceMap   & \emph{influence map}   & $b$                       & \emph{liveness bit}  \\
    \assignBuffer   & \emph{assign. buffer}  & $\pi$                     & \emph{trace}         \\
    \eventSet       & \emph{event set}       & $\StoreOf{\pi, i}$        & $\store_i$           \\
    \Program        & \emph{current program} & $\TaintOf{\pi, i}$        & $\taintMap_i$        \\
    \InitThreats    & \emph{initial program} & $\InfluenceMapOf{\pi, i}$ & $\influenceMap_i$    \\
    \Cycle          & \emph{clock cycle}     & $\CycleOf{\pi, i}$        & $\Cycle_i$           \\
                    &                        & $\IssueOf{\pi, i}$        & $b_i$                \\
    \bottomrule
  \end{tabular}
  \caption{Configuration and trace syntax.}
  \label{fig:config-syntax}
\end{figure}

\subsection{Semantics}
%
%
%
%
%

\mypara{Values} 
The set of values $\Values \eqdef \mathbb{Z} \mathrel{\uplus}
\theSet{\UnknownValue}$ consists of the disjoint union of the integers
and special value~$\UnknownValue$ which represents an irrelevant value. 
A function application that contains $\UnknownValue$ as an
argument evaluates to $\UnknownValue$.

%

\mypara{Configurations} 
The program state is represented by a \emph{configuration}
$\ConfigName \in \Config$.
Figure~\ref{fig:config-syntax} shows the components of a configuration.
A store $\store \in \Stores \eqdef \left ( \Vars \mapsto \Values \right ) $ is a
map from registers and wires to values.
A \emph{liveness map}
$\taintMap \in \TaintMaps \eqdef \left ( \Vars \mapsto \Booleans \right )$ is a
map from registers and wires to liveness bits.
A \emph{influence map}
$\influenceMap \in \InfluenceMaps \eqdef \left ( \Vars \mapsto
  \powerset(\mathbb{Z}) \right )$ is a map from registers and wires to influence
sets.
\emph{Assignment buffers} serve to model non-blocking assignments.
Let $\ProcIDs$ denote a set of process identifiers.
An assignment buffer
$\assignBuffer \in \ProcIDs \mapsto \left ( \Vars \times \Values \times
  \Booleans \times \powerset(\mathbb{Z}) \right )^\ast$ is a map from process
identifier to a sequence of variable/value/liveness-bit/influence set tuples.
An \emph{event set} $\eventSet \in \powerset(\Vars)$
is a set of variables, where we use $v \in \eventSet$
to indicate that variable $v$ has been changed in the current cycle.
Finally, $\InitThreats \in \Programs$ contains the initial program.
Intuitively, the initial program is used to activate all processes when a new
clock cycle begins.

\mypara{Evaluating Expressions}
We define an evaluation relation
$\exprTrans \in ( \Expressions \times \Stores \times \TaintMaps \times
\InfluenceMaps ) \mapsto ( \Values \times \Booleans \times \powerset(\mathbb{Z})
)$ that computes value, liveness-bit, and influence map for an expression.
We define the relation through the inference rules shown in
\cref{fig:exp-eval}. An evaluation step (below the line) can be taken,
if the preconditions (above the line) are met.
Rule [\textsc{Var}] evaluates a variable to its current value under
the store, its current liveness-bit and influence set.  A numerical
constant evaluates to itself, is dead and not influenced by any cycle. To
evaluate a function literal, we evaluate its arguments and apply
the function on the resulting values. A function value is live if any of its
arguments are, and its influence set is the union of its influences.

\begin{figure}[t]
  \figuresize
  \centering
  \begin{mathpar}
    \inferrule[\ruleName{Var}]{
    }{
      \EConfig{v}{\store}{\taintMap}{\influenceMap} \exprTrans \TaintVal{\store(v)}{\taintMap(v)}{\influenceMap(v)}
    }
    
    \inferrule[\ruleName{Const}]
    {
    }{
      \EConfig{n}{\store}{\taintMap}{\influenceMap} \exprTrans \TaintVal{n}{\TFalse}{\emptyset}
    }
    
    \inferrule[\ruleName{Fun}]
    {
      \EConfig{e_1}{\store}{\taintMap}{\influenceMap} \exprTrans \TaintVal{\Val_1}{\Taint_1}{\influenceSet_1}  \\
      \dots \\
      \EConfig{e_k}{\store}{\taintMap}{\influenceMap} \exprTrans \TaintVal{\Val_k}{\Taint_k}{\influenceSet_k} \\
      t =  \left ( \Taint_1 \lor \dots \lor \Taint_k \right ) \\
      i = \left ( \influenceSet_1 \cup \dots \cup \influenceSet_k  \right ) \\
    }{
      \EConfig{\func{e_1, \dots, e_k}}{\store}{\taintMap}{{\influenceMap}} \exprTrans \TaintVal{\func{\Val_1, \dots, \Val_k}}{t}{i}
    }
  \end{mathpar}
  \caption{Expression evaluation.}
  \label{fig:exp-eval}
\end{figure}

\mypara{Transition Relations}
We define our semantics in terms of four separate transition relations
of type $\left ( \Config \times \Labels \times \Config \right )$.
%
We now discuss
the individual relations and then describe how to
combine them into an overall transition relation~$\transRel$.

\mypara{Per-process transition~$\threadTrans$}
%
%
%
The per-process transition relation~$\threadTrans$ describes how to
step along individual processes. It is defined in \cref{fig:trans-rules}.
Rules \textsc{[Seq-Step]} and \textsc{[Par-Step]} are standard and
describe sequential and parallel composition.
Rule \textsc{[B-Asn]} reduces a blocking update $x=e$ to $\sskip$, by
first evaluating $e$ to yield a value~$v$, liveness bit~$\Taint$ and
influence set~$i$,
updating store~$\store$, liveness map~$\taintMap$ and influence
map~$\influenceMap$, and finally
adding~$x$ to the set of modified variables.
Rule~$[\textsc{NB-Asn}]$ defers a non-blocking assignment.
In order to reduce an assignment~$(\asyncAssign{x}{e})_\ID$ for
process~$\ID$ to $\sskip$, the rule evaluates expression~$e$ to
value~$v$, liveness bit~$\Taint$ and influence set $i$, and defers the
assignment by appending the tuple~$\left(x, v, \Taint, i \right)$ to
the back of $\ID$'s buffer.
We omit rules for conditionals and structural equivalence.
Structural equivalence allows transitions between trivially equivalent
programs such as $P \inPar Q$ and $Q \inPar P$.

\mypara{Non-blocking Transition~$\asyncTrans$}
Transition relation~$\asyncTrans$ applies deferred non-blocking
assignments.
It is defined by a single rule~\textsc{[NB-App]} shown in
\cref{fig:trans-rules}.
The rule first picks a tuple $ \left (x, v, \Taint, i \right )$ from
the front of the buffer of some process~$\ID$, and, like
\textsc{[B-Asn]}, updates store~$\store$,
liveness map~$\taintMap$ and influence map~$\influenceMap$, and
finally adds~$x$ to the set of updated variables.

\mypara{Continuous Transition~$\contTrans$}
Relation~$\contTrans$ specifies how to execute continuous
assignments. It is described by rule~[\textsc{C-Asn}] in
\cref{fig:trans-rules}, which reduces a continuous
assignment~$\contInter{x}{e}$ to~$\sskip$ under the condition that
some variable~$y$ occurring in~$e$ has changed, \ie 
$y \in \eventSet$. To
apply the assignment, it evaluates~$e$ to value, liveness bit and
influence set, and
updates store and liveness map and influence map. Importantly,
variable~$y$ is not removed from the set of events, \ie a single
assignment can enable several continuous assignments.

\mypara{Global Transition~$\globalTrans$}
Finally, global transition relation~$\globalTrans$ is defined by rules
\RstRule and \RstIssueRule shown in \cref{fig:trans-rules}.
\RstRule starts a new clock cycle by discarding the current program
and event set, emptying the assignment buffer, resetting the wires to
some non-deterministically chosen state (as wires only hold their value
\emph{within} a cycle), and rescheduling and activating a new set of
processes, extracted from initial program $\InitThreats$.
For a program~$P$, let~$\repeatThreads(P) \in \powerset(\Programs)$ denote the
set of processes that occur under $\repeatName$.
For a set of programs~$S$, we let $\BigPar{S}$ denote their parallel
composition.
\RstRule uses these constructs to reschedule all processes that appear
under $\repeatName$ in $\InitThreats$.
Both sources and wires are set to~$\TFalse$.
The influence map is updated by mapping all wires to the empty set, and each
source to the set containing only the current cycle.
 
\RstIssueRule performs the same step, but additionally updates the liveness map
by issuing new live bits for the source variables.
Both rules increment the cycle counter~$\Cycle$.
The rules issue a \emph{label}
$l \in \Labels \eqdef \left ( \left ( \Stores \times \TaintMaps \times
    \InfluenceMaps \times \mathbb{N} \times \theSet{\TTrue, \TFalse} \right )
  \uplus \epsilon \right )$ which is written above the arrow (all previous rules
issue the empty label~$\epsilon$).
The label contains the current store, liveness map, influence map, clock cycle,
and a bit indicating whether new live-bits have been issued.
Labels are used to construct the \textit{trace} of an execution, as we will
discuss later.

\mypara{Overall Transition~$\transRel$}
We define the overall transition relation~$\transRel \; \in \Config
\times \Labels \times \Config$ by fixing an order in which to apply
the relations.
Whenever a \emph{continuous assignment} step (relation
$\contTrans$) can be applied, that step is taken.
Whenever no continuous assignment step can be applied,
however, a \emph{per-process} step (relation $\threadTrans$) can be
applied, a $\threadTrans$ step is taken.
If no continuous assignment and process local steps can be
applied, however, an \emph{non-blocking assignment} step (relation
$\asyncTrans$) is applicable, a $\asyncTrans$ step is taken.
Finally, if neither continuous assignment, per-process, or
non-blocking steps can be applied, the program moves to a new clock
cycle by applying a \emph{global step} (relation
$\globalTrans$).
Our overall transition relation closely follows the Verilog simulation reference
model from Section 11.4 of the standard~\cite{Verilog-standard}.

\mypara{Executions and Traces}
An \emph{execution} is a finite sequence of
configurations and transition labels~$r \eqdef \Sigma_0 \Label_0
\Sigma_1 \dots \Sigma_{m-1} \Label_{m-1} \Sigma_{m}$ such that
$\Sigma_i  \overset{\Label_i}{\transRel} \Sigma_{i+1}$ for $i \in \theSet{1,\dots, m-1}$. 
We call $\Sigma_0$ \emph{initial state}
and require that all taint bits are set to~$\TFalse$, the influence map maps each
variable to the empty set, the assignment buffer is empty,
the current program is the empty program $\emp$, and the clock is set to
$0$.
The \emph{trace} of an execution is the sequence of its (non-empty)
labels.
For a trace~$\pi \eqdef (\store_0, \taintMap_0, \influenceMap_0, \Cycle_0, b_0) \dots (\store_{n-1},
\taintMap_{n-1}, \influenceMap_{n-1}  \Cycle_{n-1}, b_{n-1}) \in \Labels^\ast$ and for $i \in
\theSet{0, \dots, n-1}$ we let $\StoreOf{\pi, i} \eqdef \store_i$,
$\TaintOf{\pi, i} \eqdef \taintMap_i$, $\InfluenceMapOf{\pi, i} \eqdef
\influenceMap_i$,
$\CycleOf{\pi, i} \eqdef \Cycle_i$ and $\IssueOf{\pi, i}=b_i$, and say
the trace has length $n$. 
For a program~$P$ we use $\TracesOf{P} \in \powerset(\Labels^\ast)$ to
denote the set of its traces, \ie all traces with initial program~$P$.
%
%
%
%
%
%
\begin{figure*}[t]
  \figuresize
  \centering
  \begin{mathpar}
    \inferrule[\ruleName{Seq-Step}]
    {
      \config{\store}{\assignBuffer}{\eventSet}{\taintMap}{\statement_1}{\influenceMap}
      \threadTrans
      \config{\store'}{\assignBuffer'}{\eventSet'}{\taintMap'}{\statement_1'}{\influenceMap'}
    }{
      \config{\store}{\assignBuffer}{\eventSet}{\taintMap}{\proc{\statement_1; \statement_2}}{\influenceMap}
      \threadTrans 
      \config{\store'}{\assignBuffer'}{\eventSet'}{\taintMap'}{\proc{\statement_1'; \statement_2}}{\influenceMap'}
    }
    
    \inferrule[\ruleName{Par-Step}]
    {
      \config{\store}{\assignBuffer}{\eventSet}{\taintMap}{P}{\influenceMap} 
      \threadTrans
      \config{\store'}{\assignBuffer'}{\eventSet'}{\taintMap'}{P'}{\influenceMap'}
    }{
      \config{\store}{\assignBuffer}{\eventSet}{\taintMap}{P \inParSync Q}{\influenceMap}
      \threadTrans
      \config{\store'}{\assignBuffer'}{\eventSet'}{\taintMap'}{P' \inParSync Q}{\influenceMap'}
    }
    
    \inferrule[\ruleName{B-Asn}]
    {
      \EConfig{e}{\store}{\tau}{\influenceMap} \exprTrans \TaintVal{v}{\Taint}{\influenceSet} \\
      \store' = \upd{\store}{x}{v} \\ 
      \taintMap' = \upd{\taintMap}{x}{\Taint} \\
      \influenceMap' = \upd{\influenceMap}{x}{\influenceSet}
    }{
      \config{\store}{\assignBuffer}{\eventSet}{\taintMap}{\assign{x}{e}}{\influenceMap} 
      \threadTrans
      \config{\store'}{\assignBuffer}{\eventSet \cup \theSet{x}}{\taintMap'}{\sskip}{\influenceMap'}
    }

    \inferrule[\ruleName{NB-Asn}]
    {
      \EConfig{e}{\store}{\tau}{\influenceMap} \exprTrans \TaintVal{v}{\Taint}{\influenceSet} \\
      \assignBuffer' = \upd{\assignBuffer}{\ID}{\left( \TaintBuffAsn{x}{v,t,i} \right) \cdot q}
    }{
      \config{\store}{\upd{\assignBuffer}{\ID}{q}}{\eventSet}{\taintMap}{\left ( \asyncAssign{x}{e} \right )_\ID }{\influenceMap}
      \threadTrans
      \config{\store}{\assignBuffer'}{\eventSet}{\taintMap}{\sskip}{\influenceMap}
    }
    
    \inferrule[\ruleName{NB-App}]
    {
      \store' = \upd{\store}{x}{v} \\
      \assignBuffer' = \upd{\assignBuffer}{\ID}{q} \\
      \influenceMap' = \upd{\influenceMap}{x}{\influenceSet} \\
      \taintMap' = \upd{\taintMap}{x}{\Taint} \\
      \eventSet'=\eventSet \cup \theSet{x} \\
    }{
      \config{\store}{\upd{\assignBuffer}{\ID}{q \cdot \left( \TaintBuffAsn{x}{v,t,i} \right)}}{\eventSet}{\taintMap}{P}{\influenceMap}
      \asyncTrans 
      \config{\store'}{\assignBuffer'}{\eventSet'}{\taintMap'}{P}{\influenceMap'}
    }
    
    \inferrule[\ruleName{C-Asn}]
    {
      \EConfig{e}{\store}{\tau}{\influenceSet}  \exprTrans \TaintVal{v}{\Taint}{\influenceSet} \\
      y \in \VarsOf{e} \\
      \store' = \upd{\store}{x}{v} \\
      \taintMap' = \upd{\taintMap}{x}{\Taint} \\
      \influenceMap'= \upd{\influenceMap}{x}{\influenceSet} \\
    }{
      \config{\store}{\assignBuffer}{\eventSet \cup \theSet{y}}{\taintMap}{\contInter{x}{e}}{\influenceMap}
      \contTrans
      \config{\store'}{\assignBuffer}{\eventSet \cup \theSet{x,y}}{\taintMap'}{\sskip}{\influenceMap'}
    }
    
    \inferrule[\RstRule]
    {
      \store' \eqdef \upd{\store}{\Wires}{\ast} \\
      \taintMap' \eqdef \upd{\upd{\taintMap}{\Sources}{\TFalse}}{\Wires}{\TFalse} \\
      \influenceMap' \eqdef \upd{\upd{\influenceMap}{\Wires}{\emptyset}}{\Sources}{\theSet{\Cycle+1}} \\
      \assignBuffer' \eqdef \upd{\assignBuffer}{\ProcIDs}{\epsilon} \\
    }{
      \configClk{\store}{\assignBuffer}{\eventSet}{\taintMap}{P}{\Cycle}{\influenceMap} 
      \overset{(\store, \taintMap, \influenceMap, \Cycle,\TFalse)}{\globalTrans}
      \configClk{\store'}{\assignBuffer'}{\emptyset}{\taintMap}{\BigPar{\repeatThreads(\InitThreats)}}{\Cycle+1}{\influenceMap'}
    }
    
    \inferrule[\RstIssueRule]
    {
      \store' \eqdef \upd{\store}{\Wires}{\ast} \\
      \taintMap' \eqdef \upd{\upd{\taintMap}{\Sources}{\TTrue}}{\left( \Vars-\Sources \right)}{\TFalse} \\
      \influenceMap' \eqdef \upd{\upd{\influenceMap}{\Wires}{\emptyset}}{\Sources}{\theSet{\Cycle+1}} \\
      \assignBuffer' \eqdef \upd{\assignBuffer}{\ProcIDs}{\epsilon} \\
    }{
      \config{\store}{\assignBuffer}{\eventSet}{\taintMap}{P}{\influenceMap}
      \overset{(\store, \taintMap, \influenceMap, \Cycle, \TTrue)}{\globalTrans} 
      \configClk{\store'}{\assignBuffer'}{\emptyset} {\taintMap'}{\BigPar{\repeatThreads(\InitThreats)}}{\Cycle +1}{\influenceMap'}
    }
  \end{mathpar}
  \caption{Per-thread transition relation~$\threadTrans$, non-blocking transition relation~$\asyncTrans$, continuous transition relation~$\contTrans$, and global restart relation~$\globalTrans$.}
  \label{fig:trans-rules}
\end{figure*}



%
\section{Constant-Time Execution~\label{sec:taint}}
We now first define constant-time execution with respect to a set of
assumptions.
We then define liveness equivalence and
show that the two notions are equivalent.

\subsection{Constant-Time Execution}

\mypara{Assumptions}
For a formula $\varphi$ that ranges over two disjoint sets of
variables $\leftOf{\Vars}$ and $\rightOf{\Vars}$ and
stores $\leftOf{\store}$ and $\rightOf{\store}$ such that
$\domain \; \leftOf{\store} = \leftOf{\Vars}$ and $\domain \;
\rightOf{\store} = \rightOf{\Vars}$, we write $\leftOf{\store},
\rightOf{\store} \models \varphi$ to denote that formula~$\varphi$
holds when evaluated on $\leftOf{\store}$ and
$\rightOf{\store}$.
For some program~$P$ and a set of assumptions $\Assumptions \eqdef
\left ( \IntitEqVars, \AlwaysEqVars \right )$, we say that two traces
$\leftOf{\pi}, \rightOf{\pi} \in \TracesOf{P}$ of length~$n$
\emph{satisfy} $\Assumptions$ if
\emph{i)} for each formula~$ \varphi_I \in \IntitEqVars$, $\varphi_I$
holds initially, and
\emph{ii)} for each formula $\varphi_A \in \AlwaysEqVars$, $\varphi_A$ hold throughout, \ie
$\StoreOf{\leftOf{\pi}, 0}, \StoreOf{\rightOf{\pi},
    0} \models \varphi_I$ and 
$\StoreOf{\leftOf{\pi}, i}, \StoreOf{\rightOf{\pi},
    i} \models \varphi_A$, for $0 \leq i \leq n-1$. 
  Intuitively, pairs of traces that satisfy the assumptions 
  are ``low'' or ``input'' equivalent.
  
\mypara{Constant Time Execution}
For a program~$P$, assumptions~$\Assumptions$ and traces
$\leftOf{\pi}, \rightOf{\pi} \in \TracesOf{P}$ of length~$n$ that
satisfy~$\Assumptions$, $\leftOf{\pi}$ and $\rightOf{\pi}$ are
\emph{constant time} with respect to~$\Assumptions$, if they produce
the same influence sets for all sinks, \ie
$\InfluenceMapOf{\leftOf{\pi},i}(v)
=\InfluenceMapOf{\rightOf{\pi},i}(v)$, for~$0 \leq i \leq n-1$ and all
$v \in \Sinks$, and where two sets are equal if they contain the same
elements. A program is constant time with respect to
$\Assumptions$, if all pairs of its traces that
satisfy~$\Assumptions$ are constant time.

\subsection{Liveness Equivalence}
%
%
\mypara{$t$-Trace}
For a trace~$\pi$, we say that $\pi$ is a $t$-trace, if
$\IssueOf{\pi, t}=\TTrue$ and $\IssueOf{\pi,
  i}=\TFalse$, for $i \neq t$.

\mypara{Liveness Equivalence}
For a program~$P$, let~$\leftOf{\pi}, \rightOf{\pi} \in \TracesOf{P}$,
such that both $\leftOf{\pi}$ and $\rightOf{\pi}$ are 
of length~$n$. We say that $\leftOf{\pi}$ and
$\rightOf{\pi}$ are \emph{$t$-liveness equivalent}, if both are
$t$-traces, and 
$\TaintOf{\leftOf{\pi},i}(v) =\TaintOf{\rightOf{\pi},i}(v)$, for~$0
\leq i \leq n-1$ and all $v \in \Sinks$.
A program is $t$-liveness equivalent, with respect to a set of
assumptions~$\Assumptions$, if all pairs of $t$-traces
that satisfy~$\Assumptions$ are $t-$liveness equivalent. Finally, a
program is liveness equivalent with respect to $\Assumptions$, if it
is $t$-liveness equivalent with respect to $\Assumptions$, for
all $t$. 
%


%
\subsection{Equivalence}
We can now state our equivalence theorem.
\begin{theorem}
  \label{theorem:eq}
  For all programs~$P$ and assumptions~$A$, $P$ executes in
constant-time with respect to~$A$ if and only if it is liveness
equivalent with respect to~$A$.
\end{theorem}
\noindent
We first give a lemma which states that, if a register is $t$-live, then
$t$ is in its influence set.
\begin{lemma}
For any $t$-trace $\pi$ of length~$n$, index~$0 \leq i \leq n-1$, and variable~$v$,
if $v$ is $t-$live, \ie $\TaintOf{\pi,i}(v)=\TTrue$, then $t$ is in
$v$'s influence map, \ie $t
\in \InfluenceMapOf{\pi,i}(v)$.
\label{lemma:live-inf}
\end{lemma}
\noindent
We can now state our proof for \cref{theorem:eq}.
\begin{proof}[Proof \cref{theorem:eq}]
  \label{equiv-proof}
  The interesting direction is ``right-to-left'', \ie we want to show
that a liveness equivalent program is also constant-time.
  We prove the contrapositive, \ie if a program violates
constant-time, it must also violate liveness equivalence. For a proof by
contradiction, we assume that $P$
violates constant time execution, but satisfies liveness
equivalence. If $P$ violates constant-time execution, then there must
be a sink~$\specialOf{v}$, two trace~$\leftOf{\specialOf{\pi}},
\rightOf{\specialOf{\pi}} \in \TracesOf{P}$ that satisfy $A$, and some
index $\specialOf{i}$ such that
$\InfluenceMapOf{\leftOf{\specialOf{\pi}}, \specialOf {i}}(\specialOf
{v}) \neq \InfluenceMapOf{\rightOf{\specialOf{\pi}},
\specialOf{i}}(\specialOf{v})$, and therefore without loss of
generality, there is a cycle $\specialOf{t}$, such that $\specialOf{t}
\in \InfluenceMapOf {\leftOf{\specialOf{\pi}}, \specialOf{i}}(\specialOf{v})$ and
$\specialOf{t} \not \in \InfluenceMapOf{\rightOf{\specialOf{\pi}},
\specialOf{i}}(\specialOf{v})$. We can find two traces
$\specialOf{t}$-traces $\leftOf{\hat{\pi}}$ and $\rightOf{\hat{\pi}}$
that only differ from $\leftOf{\specialOf{\pi}}$ and
$\rightOf{\specialOf{\pi}}$ in their liveness maps. But then, since the traces are $\specialOf{t}$-liveness
  equivalent, by definition, at index~$\specialOf{i}$ both $\leftOf{\hat{\pi}}$ and
$\rightOf{\hat{\pi}}$ are $\specialOf{t}-$live, \ie
$\TaintOf{\leftOf{\hat{\pi}},\specialOf{i}}(\specialOf{v})
=\TaintOf{\rightOf{\hat{\pi}},\specialOf{i}}(\specialOf{v}) = \TTrue$
and, by \cref{lemma:live-inf},  $\specialOf{t} \in \InfluenceMapOf{\rightOf{\hat{\pi}},
\specialOf{i}}(\specialOf{v})$.
Since $\rightOf{\hat{\pi}}$ and $\rightOf{\specialOf{\pi}}$ only
differ in their liveness map, this implies
$\specialOf{t} \in \InfluenceMapOf{\rightOf{\specialOf{\pi}},
\specialOf{i}}(\specialOf{v})$, from which the contradiction follows.
\end{proof}

\section{Verifying Constant Time Execution}
\label{sec:vcgen}

In this section, we describe how \sys verifies liveness
equivalence by using standard techniques.

\mypara{Algorithm \Tool}
\label{sec:algorithm}
Given a \InterLang program $P$, a set of input/output specifications
$\SourceSink$ and a set of assumptions $\Assumptions$, \Tool checks
that $P$ executes in constant time with respect to $ \Assumptions$. 
For this, \Tool first checks for race-freedom. If a race is detected,
\Tool returns a witness describing the violation. If no race is
detected, \sys takes the following four steps:
\textbf{(1)} It builds a set of Horn clause
constraints~$hs$~\cite{Bjorner2015-sd,AndreyHornClauses} whose
solution characterizes the set of all configurations that are
reachable by the per-process product and satisfy $\Assumptions$. 
\textbf{(2)} Next, it builds a set of constraints~$cs$ whose solutions
characterize the set of liveness equivalent states.
\textbf{(3)} It then computes a solution~$Sol$ to $hs$ and checks whether the
solution satisfies~$cs$.
To find a more precise solution, the user can supply additional hints in the
form of a set of predicates which we describe later.
\textbf{(4)} If the check succeeds, $P$ executes in constant time with respect
to~$\Assumptions$, otherwise, $P$ can potentially exhibit timing variations.


\mypara{Constraint Solving}
\Tool solves the reachability constraints by using Liquid
Fixpoint~\cite{liquid-fixpoint}, which computes the \emph{strongest solution}
that can be expressed as a conjunction of elements of a set of logical formulas.
These formulas are composed of a set of \emph{base predicates}.
We use base predicates that track equalities between the liveness bits and
values of each variable between the two runs.
In addition to these base predicates, we use hints that are defined by the user.
We discuss in \cref{sec:evaluation} which predicates were used in our
benchmarks.


%

%
\section{Implementation and Evaluation}
\label{sec:evaluation}
In this section, we describe our implementation and evaluate \sys on several
open source \Verilog projects, spanning from RISC processors, to floating-point
units and crypto cores.
We find that \sys is able to show that a piece of code is not constant-time and
otherwise verify that the hardware is constant-time in a matter of seconds.
Except our processor use cases, we found the annotation burden
to be light weight---often less than 10 lines of code.
All the source code and data are available on GitHub, under an open source
license.\footnote{\url{https://iodine.programming.systems}}

\subsection{Implementation}
\sys consists of a front-end pass, which takes
annotated hardware descriptions and compiles them to \InterLang, and a
back-end that verifies the constant-time execution of these \InterLang
programs. We think this modular designs will make it easy for \sys to be extended
to support different hardware description languages beyond \Verilog
(\eg VHDL or Chisel~\cite{CHISEL}).

Our front-end extends the Icarus Verilog
parser~\cite{icarus-verilog} and consists of ~2000 lines of C++.
Since \InterLang shares many similarities with \Verilog, this pass is
relatively straightforward, however, \sys does not distinguish between
clock edges (positive or negative) and, thus, removes them during compilation.
Moreover, our prototype does not support the whole \Verilog language
(\eg we do not support assignments to multiple variables).
%

%
\sys's back-end takes a \InterLang program and, following \Cref{sec:vcgen},
generates and checks a set of verification conditions.
We implement the back-end in 4000 lines of Haskell.
Internally, this Haskell back-end generates Horn clauses and solves
them using the \textsf{liquid-fixpoint} library that wraps the \textsc{Z3}
\cite{DeMoura2008-te} SMT solver.
Our back-end outputs the generated invariants, which 
\textbf{(1)} serve as the proof of correctness when the verification succeeds, or
\textbf{(2)} helps pinpoint why verification fails.

\mypara{Tool Correctness}
The \sys implementation and Z3 SMT solver~\cite{DeMoura2008-te} are part of our
trusted computing base.
This is similar to other constant-time and information flow tools (\eg
SecVerilog~\cite{SecVerilog} and ct-verif~\cite{CTSoftware}).
As such, the formal guarantees of \sys can be undermined by implementation
bugs.
We perform several tests to catch such bugs early---in particular, we validate:
(1) our translation into \InterLang against the original \Verilog code;
(2) our translation from \InterLang into Horn clauses against our semantics;
and,
(3) the generated invariants against both the \InterLang and \Verilog code.

\subsection{Evaluation \label{sec:eval}}

Our evaluation seeks to answer three questions:
(\qone) Can \sys be easily applied to existing hardware designs?
(\qtwo) How efficient is \sys?
(\qthr) What is the annotation burden on developers?

\mypara{(\qone) Applicability}
To evaluate its applicability, we run \sys on several open
source hardware modules from GitHub and OpenCores.
We chose \Verilog programs that fit into three
categories---processors, crypto-cores, and floating-point units
(FPUs)---these have previously been shown to expose timing side
channels.
In particular, our benchmarks consist of:
\begin{CompactItemize}
\item MIPS- and RISCV-32I-based pipe-lined CPU cores with a single level memory
  hierarchy.
\item Crypto cores implementing the SHA~256 hash function and
  RSA 4096-bit encryption.
\item Two FPUs that implement core operations ($+, -, \times, \div$) according
  to the IEEE-754 standard.
\item An ALU \cite{CTALU} that implements ($+, -, \times, \ll,\dots$).
\end{CompactItemize}
%
%
In our benchmarks, following our attacker model from \Cref{ct-def}, 
we annotated all the inputs to the computation. 
%
For example, this includes the sequence of instructions for the 
benchmarks with a pipeline (\ie MIPS, RISC-V, FPU and FPU2) in 
addition to other control inputs, and all the top level \Verilog 
inputs for the rest (\ie SHA-256, ALU and RSA).
Similarly, we annotated as sinks, all the outputs of the computation.
%
In the case of benchmarks with a pipeline, this includes the output from the
last stage and other results (\eg whether the result is NaN in FPU), and all
the top level \Verilog outputs for the rest.
The modifications we had to perform to run \sys on these benchmarks were
minimal and due to parser restrictions (\eg desugaring assignments to multiple
variables into individual assignments, unrolling the code generated by the loop
inside the $\stmt{generate}$ blocks).


\begin{table}[t]
  \centering
  \figuresize
  \resizebox{\columnwidth}{!}{%
  \begin{tabular}{lrrrcr}
    \toprule

    \multirow{2}{*}{\textbf{Name}} & \multirow{2}{*}{\textbf{\#LOC}} & \multicolumn{2}{c}{\textbf{\#Assum}} & \multirow{2}{*}{\textbf{CT}} & \multirow{2}{*}{\textbf{Check} (s)} \\
    & & \textbf{\#flush} & \textbf{\#always} & & \\

    \midrule
    
    MIPS~\cite{MIPS}        & 434  & 31 & 2  & \tickYes & 1.329  \\
    RISC-V~\cite{Yarvi}     & 745  & 50 & 19 & \tickYes & 1.787  \\
    SHA-256~\cite{Sha-core} & 651  & 5  & 3  & \tickYes & 2.739  \\
    FPU~\cite{FPU}          & 1182 & 0  & 0  & \tickYes & 12.013 \\
    ALU~\cite{CTALU}        & 913  & 1  & 5  & \tickYes & 1.595  \\
    FPU2~\cite{FPU2}        & 272  & 3  & 4  & \tickNo  & 0.705  \\
    RSA~\cite{RSA}          & 870  & 4  & 0  & \tickNo  & 1.061  \\

    \midrule
    \textbf{Total} & 5067 & 94 & 33 &  -  & 21.163 \\
    \bottomrule
  \end{tabular}%
  }
  \caption{%
\textbf{\#LOC} is the number of lines of Verilog code,
\textbf{\#Assum} is the number of assumptions (excluding {\annotColor
\stmt{source}} and {\annotColor \stmt{sink}}); \textbf{flush} and
\textbf{always} are annotations of the form \initEqualName and
\alwaysEqualName respectively,
\textbf{CT} shows if the program is constant-time,
and \textbf{Check} is the time \Tool took to check the program.
All experiments were run on a Intel Core i7 processor with 16
GB RAM.
}
  \label{tab:evaluation}
\end{table}


\mypara{(\qtwo) Efficiency}
To evaluate its efficiency, we run \sys on the annotated programs.
As highlighted in \Cref{tab:evaluation}, \sys can successfully verify different
\Verilog programs of modest size (up to 1.1K lines of code) relatively
quickly (<20s).
All but the constant-time FPU finished in under 3 seconds.
Verifying the constant-time FPU took 12 seconds, despite the complexity of
IEEE-754 standard which manifests as a series of case splits in \Verilog.
We find these measurements encouraging, especially relative to the
time it takes to synthesize \Verilog---verification is orders of
magnitude smaller.

\mypara{Discovered Timing Variability}
Running \sys revealed that two of our use cases are not constant-time:
one of the FPU implementations and the RSA crypto-core.
The division module of the FPU exhibits timing variability depending on the value of 
the operands.
In particular, similar to the example from~\cref{sec:overview}, the module triggers a fast path if the operands are special
values.
 
The RSA encryption core similarly exhibited time variability.
In particular, the internal modular exponentiation algorithm performs a
Montgomery multiplication depending on the value of a source bit $e_i$:
$  \stmt{if} \; e_i = 1 \; \stmt{then} \; \overline{c} := \stmt{ModPro}(\overline{c}, \overline{m})$.
Since $e$ is a secret, this timing variability can be exploited to reveal the
secret key~\cite{kocher1996timing, brumley2005remote}.

\mypara{(\qthr) Annotation burden}
While \sys automatically discovers proofs, the user has to provide
a set of assumptions~$\Assumptions$ under which the hardware design
executes in constant time.
To evaluate the burden this places on developers, we count the
number and kinds of assumptions we had to add to each of our use
cases.
\Cref{tab:evaluation} summarizes our results: except for the CPU
cores, most of our other benchmarks required only a handful of assumptions.
Beyond declaring sinks and sources, we rely on two other kinds of annotations.
First, we find it useful to specify that the initial state of an input variable $x$
is equal in any pair of runs, \ie $\initEqual{\leftOf{x}=\rightOf{x}}$.
This assumption essentially specifies that register $x$ is flushed, 
  \ie is set to a constant value, to remove any effects of a previous execution
  from our initial state.
Second, we find it useful to specify that the state of an input variable
$x$ is equal, throughout any pair of runs, \ie
$\alwaysEqual{\leftOf{x}=\rightOf{x}}$.
This assumption is important when certain behavior is expected to be
the same in both runs.
We now describe these assumptions for our benchmarks.
\begin{CompactItemize}
\item \textbf{MIPS:}
We specify that the values of the fetched instructions, and the reset
bit are the same.
\item \textbf{RISC-V:}
In addition to the assumptions required by the MIPS core, we also specify
that both runs take the same conditional branch, and that the
type of memory access (read or write) is the same in both runs
(however, the actual values remain unrestricted).
This corresponds to the assumption that programs running on the CPU do
not branch or access memory based on secret values.
Finally, \stmt{CSR} registers must not be accessed illegally (see \cref{sec:case-studies}).
\item \textbf{ALU:}
Both runs execute the same type of operations (\eg
bitwise, arithmetic), operands have the same bit width, instructions
are valid, reset pins are the same.
\item \textbf{SHA-256 and FPU (division):}
We specify that the reset and input-ready bits are the same.
\end{CompactItemize}
In all cases, we start with no assumptions and add the assumptions
incrementally by manually investigating the constant-time
``violation'' flagged by \sys.

\mypara{Identifying Assumptions}
From our experience, the assumptions that a user needs to specify
fall into three categories.
The first are straightforward assumptions---\eg that any two runs execute the
same code.
The second class of assumptions specify that certain registers need to be
flushed, \ie they need to initially be the same (flushed) for any two runs.
To identify these, we first flush large parts of circuits, and then, in a
minimization step, we remove all unnecessary assumptions.
The last, and most challenging, are implicit invariants
on data and control---\eg the constraints on CSR registers.
\sys performs delta debugging to help pinpoint violations but, ultimately,
these assumptions require user intervention to be resolved.
Indeed, specifying these assumptions require a deep understanding of the
circuit and its intended usage.
In our experience, though, only a small fraction of assumptions fall into this
third category. 

\mypara{User Hints}
For one of our benchmarks (FPU), we needed to supply a small number of
user hints (<5) to the solver. These hints come in the form of
predicates that track additional equalities between liveness bits of
the \emph{same} run. This is required, when
the two executions can take different control paths, yet execute in
constant time. We hope to remove those hints in the future.

\subsection{Case Studies}
\label{sec:case-studies}
We now illustrate how \Tool verifies benchmarks with
challenging features and helps explicate
conditions under which a hardware design is constant-time, using
examples from our benchmarks.

\mypara{History Dependencies}
In hardware, the result of a computation often depends on inputs from
previous cycles, \ie the computation depends on execution history.
For example, when a hardware unit is in use by a previous instruction,
the CPU stalls until the unit becomes free.

The code snippet in \cref{fig:verilog-history-code} contains a
simplified version of the stalling logic from our MIPS processor
benchmark.
On line 3, register $\stmt{Stall}$ is set to $\stmt{1}$ if
instructions in the \textit{execute} and \textit{instruction decode}
stages conflict.
Its value is then used to update the state of each pipeline stage.
In this example, if the pipeline is stalled, the value of the register
$\stmt{ID\_instr}$, which corresponds to the instruction currently executing
in the \textit{instruction decode} stage, stays the same.
Otherwise, it is updated with $\stmt{IF\_instr}$---the value coming from
the \textit{instruction fetch} stage.

Without further assumptions, \Tool flags this behavior as non-constant
time, as an instruction can take different times to process, depending
on which other instructions are before it in the pipeline.
However, after adding the assumption that any two runs execute the
\emph{same sequence of instructions}, \sys is able to prove that
$\stmt{Stall}$ has the same value in any pair of traces, from which the
constant time behavior follows.
Importantly, however, we have no assumption on the state of the
registers and memory elements that the instructions use.
%
%

\begin{figure}
  \begin{lstlisting}[style=verilog-style]
always @(*) begin
  if (...)
    Stall = 1; else Stall = 0;
end
always @(posedge clk) begin
  if (Stall)
    ID_instr <= ID_instr;
  else
    ID_instr <= IF_instr;
end
  \end{lstlisting}
  \caption{Stalling in MIPS~\cite{MIPS}.}
  \label{fig:verilog-history-code}
\end{figure}

\mypara{Diverging Control Flow}
Methods for enforcing constant time execution of software often
require that any two executions take the same control flow
path~\cite{CTSoftware}.
In hardware, this assumption is too restrictive.
Consider the code snippet in \cref{fig:verilog-branch-code} taken from
our constant time FPU benchmark (the full logic is shown in
\cref{fig:verilog-branch-code-full} of the Appendix).
The first always block calculates the sign bit of
the multiplication result (\verb|sign_mul_r|), using inputs \verb|opa| and
\verb|opb|.
The FPU uses this bit in line 17 (through \verb|sign_mul_final|), to
calculate output \verb|out| in line 12.
Even though we cannot assume that all executions select the same
branches, \Tool can infer that every branch produces the same
\emph{influence sets} for the variables assigned under them.
Using this information, \sys can prove that the FPU operates in
constant-time, despite diverging control flow paths.

\begin{figure}[t]
  \begin{lstlisting}[style=verilog-style]
always @(*)
   case({opa[31], opb[31]})
        2'b0_0: sign_mul_r <= 0;
        2'b0_1: sign_mul_r <= 1;
        ...
   endcase
...
assign sign_mul_final = (sign_exe_r & ...) ?
            !sign_mul_r : sign_mul_r;
...
always @(posedge clk)
out <= { ( ... ?
            (f2i_out_sign &
            !(qnan_d | snan_d) ) :
            (((fpu_op_r3 == 3'b010)
              & ... ?
             sign_mul_final : ...))) };
  \end{lstlisting}
  \caption{Diverging control flow in FPU~\cite{FPU}.}
  \label{fig:verilog-branch-code}
\end{figure}

\mypara{Assumptions}
\sys can be used to inform software mechanisms for mitigating timing
side-channels by explicating---and verifying---conditions under
which a circuit executes in constant time.
%
 %
 Consider \Cref{fig:verilog-csr-code}, which shows the logic for updating \textit{Control
and Status Registers (CSR)} in our RISC-V benchmark.
The wire \verb|de_illegal_csr_access|, defined on
line~\ref{line:riscv-de-illegal-csr-access} is set by checking whether a
CSR instruction is executed in non-privileged mode. For this, the
circuit compares the
machine status register \verb|csr_mstatus| to the instructions status bit.
When \verb|de_illegal_csr_access| is set, the branch instruction on
line~\ref{line:riscv-interrupt-handler} traps the error and jumps to a
predefined handler code. 
In order to prove that the cycle executes in constant-time, we add an
assumption stating that CSR registers are not accessed illegally.
This assumption translates into an obligation for software mitigation
mechanisms to ensure proper use of CSR registers. 

\begin{figure}
  \begin{lstlisting}[style=verilog-style]
wire de_illegal_csr_access =$\label{line:riscv-de-illegal-csr-access}$ 
     de_valid && 
     de_inst`opcode == `SYSTEM && 
     de_inst`funct3 != `PRIV &&
     ( csr_mstatus`PRV < de_inst[29:28] ||
       ... );
always @(posedge clk) begin
   if (de_illegal_csr_access) begin$\label{line:riscv-interrupt-handler}$
      ex_restart <= 1;
      ex_next_pc <= ...;
   end
end
  \end{lstlisting}
  \caption{Update of CSRs in RISC-V~\cite{Yarvi}.}
  \label{fig:verilog-csr-code}
\end{figure}
 

%
\section{Limitations and Future Work}
\label{sec:limitations}
%
We discuss some of  \sys{}'s limitations.

\mypara{Clocks and Assumptions}
For example, \sys presupposes a single fixed-cycle clock and thus does not allow for
checking arbitrary \Verilog programs.
%
%
We leave an extension to multiple clocks as future work.
%
%
Similarly, \sys requires users to add assumptions by hand in somewhat ad-hoc
trial-and-error fashion.
For large circuits this could prove extremely difficult and potentially lead to
errors where erroneous assumptions may lead \sys to falsely
mark a variable time circuit as constant-time.
%
We leave the inference and validation of assumptions to future work.


\mypara{Scale}
We evaluate \sys on relatively small sized (500-1000 lines) hardware designs.
We did not (yet) evaluate the tool on larger circuits, such as modern
processors with advanced features like  a memory hierarchy, and out-of-order
and transient-execution.
In principle, these features boil down to the same primitives (always
blocks and assignments) that \sys already handles. But, we anticipate
that scaling will require further changes to \sys, for instance,
finding per-module invariants rather than the naive in-lining currently
performed by \sys.
We leave the evaluation to larger systems to future work.



\section{Related Work}
\label{sec:related}
%

\mypara{Constant-Time Software}
Almeida et al.~\cite{CTSoftware} verify 
constant-time execution of cryptographic
libraries for LLVM.
Their notion of constant-time execution 
is based on a \emph{leakage model}.
This choice allows them to be flexible 
enough to capture various properties 
like (timing) variability in memory 
access patterns and improper use of 
timing sensitive 
instructions like \verb|DIV|.
Unfortunately, their notion of 
constant-time is too restrictive 
for our setting, as it requires 
the control flow path of any
two runs to be the same.
This would, for example, incorrectly 
flag our FPU multiplier as variable-time.
Like \sys, their tool \textsf{ct-verif} 
employs a product construction that 
use the fact that loops can often be 
completely unrolled in cryptographic 
code, whereas we rely on race freedom.

Barthe et al.~\cite{Barthe14} build 
on the CompCert compiler~\cite{CompCert06} 
to enforce constant time execution through 
an information flow type system.
%

Reparaz et al.~\cite{dudect} present a method for
discovering timing variability in existing systems through a black-box
approach, based on statistical measurements.

All of these approaches address constant-time 
execution in software and 
do not translate to the hardware setting
(see \cref{sec:overview}).

\mypara{Self-Composition and Product Programs}
Barthe et al.~\cite{Barthe04} introduce the 
notion of self composition to verify 
information flow. 
%
%
Terauchi and Aiken generalize this construction 
to arbitrary 2-safety properties~\cite{Terauchi05}, 
\ie properties that relate two runs, and 
Clarkson and Schneider~\cite{hyperprop} 
generalize to multiple runs.
Barthe et al.~\cite{Barthe-Product} introduce 
product programs that, instead of conjoining 
copies sequentially, compose copies in lock-step; this
was later used in other tools like \textsf{ct-verif}.
This technique is further developed in~\cite{Dillig2016}, 
which presents 
an extension of Hoare logic to 
hyper-properties that computes lock-step 
compositions on demand, per Hoare-triple.

\mypara{Information Flow Safety and Side Channels}
There are many techniques for proving information flow safety (\eg
non-interference) in both hardware and software.
Kwon et al.~\cite{Harris17} prove information flow safety of
hardware for policies that allow explicit declassification and are
expressed over streams of input data.
They construct relational invariants by using propositional
interpolation and implicitly build a full self-composition; by
contrast, we leverage race-freedom to create a per-thread
product which contains only a subset of behaviors.
 
SecVerilog~\cite{SecVerilog} proves timing-sensitive non-interference for
circuits implemented in an extension of \Verilog that uses value-dependent
information flow types. 
Caisson~\cite{li2011caisson} is a hardware description language that
uses information flow types to ensure that generated circuits are secure.
GLIFT~\cite{tiwari2009complete, Kasnter11} tracks the flow of information at the
gate level to eliminate explicit and covert channels.
All these approaches have been used to implement information flow secure hardware that do not suffer from
(timing) side-channels.

\Tool focuses on clock-precise constant-time execution,
not information flow. The two properties are related,
but information flow safety does not imply constant-time 
execution nor the converse (see \cref{sec:ifc} for details).
Moreover, SecVerilog, Caisson, and GLIFT take 
a language-design approach whereas we take an 
analysis-centric view that is more suitable 
for verifying \emph{existing} hardware designs.
Thus, we see our work as largely complementary.
Indeed, it may be useful to use \Tool alongside 
these HDLs to verify constant-time execution for 
parts of the hardware that handle secret data only, 
and are thus not checked for timing variability, 
thereby extending their attacker model.


\mypara{Combining Hardware \& Software Mitigations}
HyperFlow~\cite{ferraiuolo2018hyperflow} and
GhostRider~\cite{liu2015ghostrider}, take 
hardware/software co-design approach to 
eliminating timing channels.
Zhang et al.~\cite{MyersPLDI2012} present a method for mitigating timing
side-channels in software and give conditions on hardware that ensure the
validity of mitigations is preserved. Instead of eliminating timing flows all
together, they specify quantitative bounds on leakage and offers
primitives to mitigate timing leaks through padding.
Many other tools~\cite{Koepf2013, rodrigues2016sparse, almeida2013formal,
tis-ct, ctgrind} automatically quantify leakage through timing and 
cache side-channels.
Our approach is complementary and focuses on clock-precise analysis 
of existing hardware. However, the explicit assumptions that \sys 
needs to verify constant-time behavior can be used to inform 
software mitigation techniques.


\section*{Acknowledgements}

We sincerely thank the anonymous reviewers (including the drive-by reviewers!)
and our shepherd Stephen Checkoway for their suggestions and their insightful
and detailed comments.
This work was supported in part by gifts from Fujitsu and Cisco, by the
National Science Foundation under Grant Number CNS-1514435, by ONR Grant
N000141512750, and by the CONIX Research Center, one of six centers in JUMP, a
Semiconductor Research Corporation (SRC) program sponsored by DARPA.

\bibliographystyle{plain}
\bibliography{main}

\begin{thebibliography}{10}

\bibitem{CTALU}
\href{https://github.com/scarv/xcrypto-ref}{https://github.com/scarv/xcrypto-ref}.

\bibitem{arm-a64-isa}
{ARM} {A64} instruction set architecture.
\newblock
  \href{https://static.docs.arm.com/ddi0596/a/DDI_0596_ARM_a64_instruction_set_architecture.pdf}{https://static.docs.arm.com}.

\bibitem{FPU2}
\href{https://github.com/dawsonjon/fpu}{https://github.com/dawsonjon/fpu}.

\bibitem{RSA}
\href{https://github.com/fatestudio/RSA4096}{https://github.com/fatestudio/RSA4096}.

\bibitem{MIPS}
\href{https://github.com/gokhankici/iodine/tree/master/benchmarks/472-mips-pipelined}{https://github.com/gokhankici/iodine}.

\bibitem{FPU}
\href{https://github.com/monajalal/fpga_mc/tree/master/fpu}{https://github.com/monajalal/fpga_mc/tree/master/fpu}.

\bibitem{Yarvi}
\href{https://github.com/tommythorn/yarvi}{https://github.com/tommythorn/yarvi}.

\bibitem{Sha-core}
\href{https://opencores.org/project/sha_core}{https://opencores.org/project/sha_core}.

\bibitem{icarus-verilog}
Icarus verilog.
\newblock \href{http://iverilog.icarus.com/}{http://iverilog.icarus.com/}.

\bibitem{liquid-fixpoint}
Liquid fixpoint.
\newblock
  \href{https://github.com/ucsd-progsys/liquid-fixpoint}{https://github.com/ucsd-progsys}.

\bibitem{tis-ct}
{TIS-CT}.
\newblock
  \href{http://trust-in-soft.com/tis-ct/}{http://trust-in-soft.com/tis-ct/}.

\bibitem{Verilog-standard}
{\em IEEE Standard for Verilog Hardware Description Language.}
\newblock IEEE Std 1364-2005, 2005.

\bibitem{almeida2013formal}
J~Bacelar Almeida, Manuel Barbosa, Jorge~S Pinto, and B{\'a}rbara Vieira.
\newblock Formal verification of side-channel countermeasures using
  self-composition.
\newblock In {\em Science of Computer Programming}, 2013.

\bibitem{jasmin}
Jos{\'e}~Bacelar Almeida, Manuel Barbosa, Gilles Barthe, Arthur Blot, Benjamin
  Gr{\'e}goire, Vincent Laporte, Tiago Oliveira, Hugo Pacheco, Benedikt
  Schmidt, and Pierre-Yves Strub.
\newblock Jasmin: High-assurance and high-speed cryptography.
\newblock In {\em CCS}, 2017.

\bibitem{verify-s2n-mee-cbc}
Jos{\'e}~Bacelar Almeida, Manuel Barbosa, Gilles Barthe, and Fran{\c{c}}ois
  Dupressoir.
\newblock Verifiable side-channel security of cryptographic implementations:
  Constant-time mee-cbc.
\newblock In {\em FSE}, 2016.

\bibitem{CTSoftware}
Jos{\'e}~Bacelar Almeida, Manuel Barbosa, Gilles Barthe, Fran{\c c}ois
  Dupressoir, and Michael Emmi.
\newblock Verifying constant-time implementations.
\newblock In {\em {USENIX} Security}, 2016.

\bibitem{andrysco2015}
Marc Andrysco, David Kohlbrenner, Keaton Mowery, Ranjit Jhala, Sorin Lerner,
  and Hovav Shacham.
\newblock On subnormal floating point and abnormal timing.
\newblock In {\em S\&P}, 2015.

\bibitem{Andrysco18}
Marc Andrysco, Andres Noetzli, Fraser Brown, Ranjit Jhala, and Deian Stefan.
\newblock Towards verified, constant-time floating point operations.
\newblock In {\em CCS}, 2018.

\bibitem{CHISEL}
Jonathan Bachrach, Huy Vo, Brian~C. Richards, Yunsup Lee, Andrew Waterman,
  Rimas Avizienis, John Wawrzynek, and Krste Asanovic.
\newblock Chisel: constructing hardware in a scala embedded language.
\newblock In {\em DAC}, 2012.

\bibitem{Barthe14}
Gilles Barthe, Gustavo Betarte, Juan~Diego Campo, Carlos~Daniel Luna, and David
  Pichardie.
\newblock Systemlevel non-interference for constant-time cryptography.
\newblock In {\em CCS}, 2014.

\bibitem{Barthe-Product}
Gilles Barthe, Juan~Manuel Crespo, and Cesar Kunz.
\newblock Relational verification using product programs.
\newblock In {\em FM}, 2011.

\bibitem{Barthe04}
Gilles Barthe, Pedro~R. D'Argenio, and Tamara Rezk.
\newblock Secure information flow by self-composition.
\newblock In {\em CSF}, 2004.

\bibitem{poly1305}
Daniel~J. Bernstein.
\newblock The poly1305-aes message-authentication code.
\newblock In {\em Fast Software Encryption}, 2005.

\bibitem{donnacurve}
Daniel~J. Bernstein.
\newblock Curve25519: New diffie-hellman speed records.
\newblock In {\em Public Key Cryptography}, 2006.

\bibitem{salsa20}
Daniel~J Bernstein.
\newblock The salsa20 family of stream ciphers.
\newblock In {\em New stream cipher designs}. Springer, 2008.

\bibitem{Bjorner2015-sd}
Nikolaj Bj{\o}rner, Arie Gurfinkel, Ken McMillan, and Andrey Rybalchenko.
\newblock Horn clause solvers for program verification.
\newblock In {\em Fields of Logic and Computation}. 2015.

\bibitem{brumley2005remote}
David Brumley and Dan Boneh.
\newblock Remote timing attacks are practical.
\newblock {\em Computer Networks}, 2005.

\bibitem{hyperprop}
Michael~R. Clarkson and Fred~B. Schneider.
\newblock Hyperproperties.
\newblock {\em Journal of Computer Security}, 2010.

\bibitem{DeMoura2008-te}
Leonardo de~Moura and Nikolaj Bj{\o}rner.
\newblock Z3: An efficient {SMT} solver.
\newblock In {\em TACAS}, 2008.

\bibitem{Koepf2013}
Goran Doychev, Dominik Feld, Boris K{\"o}pf, Laurent Mauborgne, and Jan
  Reineke.
\newblock Cacheaudit: A tool for the static analysis of cache side channels.
\newblock In {\em {USENIX} Security}, 2013.

\bibitem{ferraiuolo2018hyperflow}
Andrew Ferraiuolo, Mark Zhao, Andrew~C Myers, and G~Edward Suh.
\newblock Hyperflow: A processor architecture for nonmalleable, timing-safe
  information flow security.
\newblock In {\em SIGSAC}, 2018.

\bibitem{gordon95}
Michael J.~C. Gordon.
\newblock The semantic challenge of verilog hdl.
\newblock In {\em LICS}, 1995.

\bibitem{AndreyHornClauses}
Sergey Grebenshchikov, Nuno~P. Lopes, Corneliu Popeea, and Andrey Rybalchenko.
\newblock Synthesizing software verifiers from proof rules.
\newblock In {\em PLDI}, 2012.

\bibitem{spectre}
Paul Kocher, Daniel Genkin, Daniel Gruss, Werner Haas, Mike Hamburg, Moritz
  Lipp, Stefan Mangard, Thomas Prescher, Michael Schwarz, and Yuval Yarom.
\newblock Spectre attacks: Exploiting speculative execution.
\newblock {\em CoRR}, 2018.

\bibitem{kocher1996timing}
Paul~C Kocher.
\newblock Timing attacks on implementations of {Diffie-Hellman}, {RSA}, {DSS},
  and other systems.
\newblock In {\em CRYPTO}, 1996.

\bibitem{kohlbrenner2017}
David Kohlbrenner and Hovav Shacham.
\newblock On the effectiveness of mitigations against floating-point timing
  channels.
\newblock In {\em {USENIX} Security}, 2017.

\bibitem{Harris17}
Hyoukjun Kwon, William Harris, and Hadi Esameilzadeh.
\newblock Proving flow security of sequential logic via
  automatically-synthesized relational invariants.
\newblock In {\em CSF}, 2017.

\bibitem{ctgrind}
Adam Langley.
\newblock {ctgrind}: Checking that functions are constant time with valgrind.
\newblock
  \href{https://github.com/agl/ctgrind/}{https://github.com/agl/ctgrind/}.

\bibitem{CompCert06}
Xavier Leroy.
\newblock Formal certification of a compiler back-end, or: programming a
  compiler with a proof assistant.
\newblock In {\em POPL}, 2006.

\bibitem{li2011caisson}
Xun Li, Mohit Tiwari, Jason~K Oberg, Vineeth Kashyap, Frederic~T Chong, Timothy
  Sherwood, and Ben Hardekopf.
\newblock Caisson: a hardware description language for secure information flow.
\newblock In {\em PLDI}, 2011.

\bibitem{linux-on-arm}
{Linux on ARM}.
\newblock {ARM64} prepping {ARM} v8.4 features, {KPTI} improvements for {Linux}
  4.17.
\newblock \url{https://www.linux-arm.info/}.

\bibitem{Lipp2018meltdown}
Moritz Lipp, Michael Schwarz, Daniel Gruss, Thomas Prescher, Werner Haas,
  Anders Fogh, Jann Horn, Stefan Mangard, Paul Kocher, Daniel Genkin, Yuval
  Yarom, and Mike Hamburg.
\newblock Meltdown: Reading kernel memory from user space.
\newblock In {\em USENIX Security}, 2018.

\bibitem{liu2015ghostrider}
Chang Liu, Austin Harris, Martin Maas, Michael Hicks, Mohit Tiwari, and Elaine
  Shi.
\newblock Ghostrider: A hardware-software system for memory trace oblivious
  computation.
\newblock {\em SIGPLAN Notices}, 2015.

\bibitem{magazinius2010fly}
Jonas Magazinius, Alejandro Russo, and Andrei Sabelfeld.
\newblock On-the-fly inlining of dynamic security monitors.
\newblock In {\em IFIP}, 2010.

\bibitem{OwickiGries}
Susan Owicki and David Gries.
\newblock Verifying properties of parallel programs: an axiomatic approach.
\newblock {\em Communicationsof the {ACM}}, 1976.

\bibitem{rane2016}
Ashay Rane, Calvin Lin, and Mohit Tiwari.
\newblock Secure, precise, and fast floating-point operations on x86
  processors.
\newblock In {\em {USENIX} Security}, 2016.

\bibitem{dudect}
Oscar Reparaz, Joseph Balasch, and Ingrid Verbauwhede.
\newblock Dude, is my code constant time?
\newblock In {\em DATE}, 2017.

\bibitem{rodrigues2016sparse}
Bruno Rodrigues, Fernando~Magno Quint{\~a}o~Pereira, and Diego~F Aranha.
\newblock Sparse representation of implicit flows with applications to
  side-channel detection.
\newblock In {\em CCC}, 2016.

\bibitem{Dillig2016}
Marcelo Sousa and Isil Dillig.
\newblock Cartesian hoare logic for verifying k-safety properties.
\newblock In {\em PLDI}, 2016.

\bibitem{Terauchi05}
Tachio Terauchi and Alex Aiken.
\newblock Secure information flow as a safety problem.
\newblock In {\em SAS}, 2005.

\bibitem{Kasnter11}
Mohit Tiwari, Jason~K Oberg, Xun Li, Jonathan Valamehr, Timothy Levin, Ben
  Hardekopf, Ryan Kastner, Frederic~T. Chong, and Timothy Sherwood.
\newblock Crafting a usable microkernel, processor, and i/o system with strict
  and provable information flow security.
\newblock In {\em ISCA}, 2011.

\bibitem{tiwari2009complete}
Mohit Tiwari, Hassan~MG Wassel, Bita Mazloom, Shashidhar Mysore, Frederic~T
  Chong, and Timothy Sherwood.
\newblock Complete information flow tracking from the gates up.
\newblock In {\em Sigplan Notices}, 2009.

\bibitem{watt:2019:ctwasm}
Conrad Watt, John Renner, Natalie Popescu, Sunjay Cauligi, and Deian Stefan.
\newblock Ct-wasm: Type-driven secure cryptography for the web ecosystem.
\newblock 2019.

\bibitem{MyersPLDI2012}
Danfeng Zhang, Aslan Askarov, and Andrew~C. Myers.
\newblock Language-based control and mitigation of timing channels.
\newblock In {\em PLDI}, 2012.

\bibitem{SecVerilog}
Danfeng Zhang, Yao Wang, G.~Edward Suh, and Andrew~C. Myers.
\newblock A hardware design language for timing-sensitive information-flow
  security.
\newblock In {\em ASPLOS}, 2015.

\end{thebibliography}

\appendix
\section{Appendix}
\label{sec:appendix}

\subsection{Comparison to Information Flow}
\label{sec:ifc}

In this section, we discuss the relationship between constant time execution and
information flow checking.
Information flow safety (IFS) and constant time execution (CTE) are
\emph{incomparable}, \ie IFS does not imply CTE, and vice versa.
We illustrate this using two examples: one is information flow safe but does not
execute in constant time and one executes in constant time but is not
information flow safe.
 
Figure~\ref{fig:sec-verilog-non-ct} contains example program~$\exNum{2}$ which
is information flow safe but not constant time.
The example contains three registers that are typed high as indicated by the
annotation~${\annotColor H}$, and one register that is typed low as indicated by
the annotation~${\annotColor L}$.
The program is information flow safe, as there are no flows from high to low.
Indeed, SecVerilog~\cite{SecVerilog} type checks this program.

This program, however, is not constant time when
$\leftOf{\stmt{slow}} \neq \rightOf{\stmt{slow}}$.
This does not mean that the program leaks high data to low sinks---indeed it
does not.
Instead, what this means is that the high computation takes a variable amount
of time dependent on the secret input values.
In cases like crypto cores where the attacker has a stop watch and can measure
the duration of the sensitive computation, it's not enough to be information
flow safe: we must ensure the core is constant-time.

\begin{figure}[t]
  \begin{lstlisting}[style=verilog-style,escapeinside={(*}{*)}]
// source(in_low); source(in_high);
// sink(out_low); sink(out_high);
module test(input  (*\theSet{{\annotColor{L}}}*) clk,
            input  (*\theSet{{\annotColor{L}}}*) in_low,
            input  (*\theSet{{\annotColor{H}}}*) in_high,
            output (*\theSet{{\annotColor{L}}}*) out_low,
            output (*\theSet{{\annotColor{H}}}*) out_high);
  reg (*\theSet{{\annotColor{H}}}*) flp_res;
  reg (*\theSet{{\annotColor{H}}}*) slow;
  reg (*\theSet{{\annotColor{L}}}*) out_low;
  reg (*\theSet{{\annotColor{H}}}*) out_high;
  always @(posedge clk) begin
    out_low <= in_low;
    flp_res <= in_hi;
    if (slow)
      out_hi <= flp_res;
    else
      out_hi <= in_hi;
  end
endmodule
  \end{lstlisting}
  \caption{\exNum{2}: Non-constant time but
    info-flow safe.}
  \label{fig:sec-verilog-non-ct}
\end{figure}

Next, consider Figure~\ref{fig:sec-verilog-ct} that contains
program~$\exNum{3}$ which executes in constant time but is
not information flow safe.
$\exNum{3}$ violates information flow safety by assigning high
input~$\stmt{sec}$ to low output~$\stmt{out}$.
The example however executes constant time with source~$\stmt{in}$ and
sink~$\stmt{out}$ under the assumption that~$+$ does not contain asynchronous
assignments.

\begin{figure}[t]
  \begin{lstlisting}[style=verilog-style,escapeinside={(*}{*)}]
// source(in); sink(out);
// (*\alwaysEqual{\stmt{\leftOf{\stmt{slow}}=\rightOf{\stmt{slow}}}};*)
reg (*\theSet{{\annotColor{L}}}*) in;
reg (*\theSet{{\annotColor{L}}}*) out;
reg (*\theSet{{\annotColor{H}}}*) sec;
always @(posedge clk) begin
  out <= in + sec;
end
  \end{lstlisting}
  \caption{\exNum{3}: Constant time but not info-flow safe.}
  \label{fig:sec-verilog-ct}
\end{figure}

\subsection{Translation}
\label{sec:syntax-trans}

In \Cref{fig:inter-lang}, we define a relation~$\InterTrans$ that translates
\Verilog programs into \InterLang programs.
%
%
The relation is given in terms of inference rules where a transition step in the
rule's conclusion (below the line) is applicable only if all its preconditions
(above the line) are met.
Both $\alwaysName$- and $\assignNameHL$-blocks are translated into threads that
are executed at every clock tick using $\SyncRepeatName$.
Each process is given a unique id.
Our translation does not distinguish between $\mathsf{posedge}$ and
$\mathsf{negedge}$ events thereby relaxing the semantics by allowing them to
occur in any order.
$\assignNameHL$ blocks are transformed into threads executing a continuous
assignment.
Blocking and non-blocking assignments remain unchanged.

\begin{figure}[t]
  \figuresize
  \centering
  \begin{mathpar}
    \inferrule{
      P \InterTrans P' \\
      Q \InterTrans Q' \\
    }{
      P\cdot Q \InterTrans P' \inParSync Q' 
    }
    
    \inferrule{
      \statement_1 \InterTrans \statement_1' \\
      \dots \\
      \statement_n \InterTrans \statement_n'\\
    }{
      \beginEnd{\statement_1; \dots; \statement_n;} \InterTrans \statement_1'; \dots ; \statement_n'
    }
    
    \inferrule{
      \statement \InterTrans \statement ' \\
      \ID \; \text{fresh} \\
    }{
      \always{\_ } \; \statement \InterTrans \srepeat{\proc{\statement '}_\ID}
    }

    \inferrule{
      \ID \; \text{fresh}
    }{
      \assignNameHL \ \assign{v}{e} \InterTrans \srepeat{\proc{\contInter{v}{e}}_\ID}
    }
    
    \inferrule{
      \statement_1 \InterTrans \statement_1' \\
      \statement_2 \InterTrans \statement_2' \\
    }{
      \ifThenElse{e}{\statement_1}{\statement_2}  \;\; \InterTrans \;\;  \ite{e}{\statement_1'}{\statement_2'}
    }
  \end{mathpar}
  \caption{Translation from \Verilog to \InterLang.}
  \label{fig:inter-lang}
\end{figure}

% ...

\begin{figure}[t]
  \begin{lstlisting}[style=verilog-style]
always @(*)
   case({opa[31], opb[31]})
        2'b0_0:  sign_mul_r <= 0;
        2'b0_1:  sign_mul_r <= 1;
        ...
   endcase
assign sign_mul_final =
  (sign_exe_r & 
   ((opa_00 & opb_inf) | 
    (opb_00 & opa_inf))) ? 
   !sign_mul_r : sign_mul_r;
always @(posedge clk)
  out <= { 
   (((fpu_op_r3 == 3'b101) & out_d_00) ?
    (f2i_out_sign & !(qnan_d | snan_d)) :
    (((fpu_op_r3 == 3'b010) & 
      !(snan_d | qnan_d)) ?
     sign_mul_final :
     (((fpu_op_r3 == 3'b011) & 
       !(snan_d | qnan_d)) ? sign_div_final : 
      ((snan_d | qnan_d | ind_d) ?
       nan_sign_d :
       (output_zero_fasu ? 
        result_zero_sign_d : 
        sign_fasu_r))))) ,
   ((mul_inf | div_inf | 
     (inf_d & (fpu_op_r3 != 3'b011) & 
      (fpu_op_r3 != 3'b101)) |
     snan_d | qnan_d) &
    fpu_op_r3 != 3'b100 ? out_fixed : out_d) };
  \end{lstlisting}
  \caption{Example diverging computation in \cite{FPU}}
  \label{fig:verilog-branch-code-full}
\end{figure}


\end{document}
